\documentclass{mscs}
\usepackage{boxedminipage}
\usepackage{proof}
\usepackage{amssymb}
\usepackage{amsmath}
\usepackage{amsfonts}
\usepackage{color}
\usepackage{times}
\usepackage[authoryear]{natbib}
\let\cite=\citep
\let\citeN=\citet
\newcommand{\eq}{\approx}

\newcommand{\kw}[1]{\mathsf{#1}}
\newcommand{\kbool}{\kw{bool}}

\newcommand{\kreal}{\kw{real}}
\newcommand{\kint}{\kw{int}}
\newcommand{\kcolor}{\kw{color}}
\newcommand{\kcount}{\kw{count}}
\newcommand{\ksum}{\kw{sum}}
\newcommand{\kaverage}{\kw{average}}
\newcommand{\kfin}{\kw{fin}}
\newcommand{\ktrue}{\kw{true}}
\newcommand{\kfalse}{\kw{false}}
\newcommand{\kif}{\kw{if}}
\newcommand{\kthen}{\kw{then}}
\newcommand{\kelse}{\kw{else}}
\newcommand{\klet}{\kw{let}}
\newcommand{\kin}{\kw{in}}
\newcommand{\kcond}{\kw{cond}}

\newcommand{\dc}{\kw{dc}}
\renewcommand{\tilde}{\text{\texttt{\~}}}
\renewcommand{\hat}[1]{\widehat{#1}}

\newcommand{\boolTy}{\kbool}
\newcommand{\intTy}{\kint}
\newcommand{\colorTy}{\kcolor}

\newcommand{\Int}{\mathbb{Z}}

\newcommand{\Bool}{\mathbb{B}}

\renewcommand{\AA}[1]{\widehat{\mathcal{A}}\SB{#1}}
\newcommand{\PP}[1]{\mathcal{P}\SB{#1}}
\newcommand{\AAz}[1]{\mathcal{A}\SB{#1}}

\newcommand{\setof}[1]{\{#1\}}
\newcommand{\T}[1]{\mathcal{T}\SB{#1}}
\newcommand{\E}[1]{\mathcal{E}\SB{#1}}
\newcommand{\SB}[1]{[\![#1]\!]}

\newcommand{\ifthenelse}[3]{\kif~#1~\kthen~#2~\kelse~#3}
\newcommand{\letin}[2]{\klet~ #1~\kin~#2}

\newcommand{\andd}{\wedge}
\newcommand{\orr}{\vee}

\newcommand{\nott}{\neg}

\newcommand{\wf}[3]{#1 \vdash #2:#3}

\newcommand{\labelSec}[1]{\label{sec:#1}}
\newcommand{\refSec}[1]{Section~\ref{sec:#1}}
\newcommand{\labelFig}[1]{\label{fig:#1}}
\newcommand{\refFig}[1]{Figure~\ref{fig:#1}}

\newcommand{\labelThm}[1]{\label{thm:#1}}
\newcommand{\refThm}[1]{Theorem~\ref{thm:#1}}
\newcommand{\labelLem}[1]{\label{lem:#1}}
\newcommand{\refLem}[1]{Lemma~\ref{lem:#1}}

\newcommand{\labelEx}[1]{\label{ex:#1}}
\newcommand{\refEx}[1]{Example~\ref{ex:#1}}

\newcommand{\with}{\mathrel{\&}}
\newcommand{\Color}{\mathsf{Color}}

\newcommand{\dom}{\mathrm{dom}}
\newcommand{\wfeff}[4]{\wf{#1}{#2}{#3 \with #4}}
\renewcommand{\tilde}{\text{\texttt{\~}}}
\renewcommand{\hat}[1]{\widehat{#1}}
\newcommand{\hop}[1]{\mathrel{\widehat{#1}}}
\newcommand{\htau}{\hat{\tau}}

\newcommand{\hg}{\hat{\gamma}}
\newcommand{\hG}{\hat{\Gamma}}

\newcommand{\enriches}{\gtrsim}

\newcommand{\meet}{\sqcap}

\newtheorem{lemma}{Lemma}[section]
\newtheorem{theorem}{Theorem}[section]
\newtheorem{proposition}{Proposition}[section]
\newtheorem{definition}{Definition}[section]
\newtheorem{remark}{Remark}[section]
\newtheorem{example}{Example}[section]

\begin{document}
\title{Provenance as Dependency Analysis\footnote{This paper is a revised and extended version of~\cite{cheney07dbpl}.
}} 
\author[J. Cheney and A. Ahmed and U. A. Acar]{
James Cheney$^1$
\thanks{Cheney is supported by EPSRC grant GR/S63205/01 and a Royal
  Society University Research Fellowship.}  
and 
Amal Ahmed$^2$
and 
Umut A. Acar$^3$
\thanks{Acar is
  supported by gifts from Intel and Microsoft Research.}\\
$^1$University of Edinburgh
\addressbreak
$^2$Indiana University
\addressbreak
$^3$Max-Planck Institute for Software Systems}

\maketitle

\begin{abstract}
  Provenance is information recording the source, derivation, or
  history of some information.  Provenance tracking has been studied
  in a variety of settings, particularly database management systems;
  however, although many candidate definitions of provenance have been
  proposed, the mathematical or semantic foundations of data
  provenance have received comparatively little attention.  In this
  article, we argue that dependency analysis techniques familiar from
  program analysis and program slicing provide a formal foundation for
  forms of provenance that are intended to show how (part of) the
  output of a query \emph{depends} on (parts of) its input.  We introduce a
  semantic characterization of such \emph{dependency provenance} for a
  core database query language, show that minimal dependency
  provenance is not computable, and provide dynamic and static
  approximation techniques.  We also discuss preliminary
  implementation experience with using dependency provenance to
  compute data slices, or summaries of the parts of the input
  relevant to a given part of the output.
\end{abstract}

%--------------------------------------------------------------------%
\section{Introduction}\labelSec{intro}
%--------------------------------------------------------------------%

\emph{Provenance} is information about the origin, ownership,
influences upon, or other historical or contextual
information about an object.  Such information has many applications,
including evaluating integrity or authenticity claims, detecting and
repairing errors, and memoizing and caching the results of
computations such as scientific
workflows~\cite{lynch00clir,bose05cs,DBLP:journals/sigmod/SimmhanPG05}.
Provenance is particularly important in scientific computation and
recordkeeping, since it is considered essential for ensuring the
repeatability of experiments and judging the scientific value of their
results~\cite{buneman08pods}.

Most computer systems provide simple forms of provenance, such as the
timestamp and ownership metadata in file systems, system logs, and
version control systems.  Richer provenance tracking techniques have
been studied in a variety of settings, including
databases~\cite{DBLP:journals/tods/CuiWW00,buneman01icdt,buneman08tods,foster08pods,green07pods},
file systems~\cite{muniswamy-reddy06usenix}, and scientific
workflows~\cite{bose05cs,DBLP:journals/sigmod/SimmhanPG05}.  Although
a wide variety of design points have been explored, there is
relatively little understanding of the relationships among techniques
or of the design considerations that should be taken into account when
developing or evaluating an approach to provenance.  The mathematical
or semantic foundations of data provenance have received comparatively
little attention.  Most prior approaches have invoked intuitive
concepts such as \emph{contribution}, \emph{influence}, and
\emph{relevance} as motivation for their definitions of provenance.
These intuitions suggest definitions that appear adequate for simple
(e.g. conjunctive) relational queries, but are difficult to extend to
handle more complex queries involving subqueries, negation, grouping,
or aggregation.

However, these intuitions have also motivated rigorous approaches to
seemingly quite different problems, such as aiding debugging via
\emph{program slicing}~\cite{biswas97phd,field98ist,weiser81icse},
supporting efficient memoization and
caching~\cite{abadi96icfp,AcarBlHa03}, and improving program security
using information flow analysis~\cite{sabelfeld03sac}.  As
\citeN{abadi99popl} have argued, slicing, information flow, and
several other program analysis techniques can all be understood in
terms of dependence.
In this article, we argue that these {dependency analysis} and
slicing techniques familiar from programming languages provide
a suitable foundation for an interesting class of provenance
techniques.

\begin{figure}[tb]
(a)
\[
\begin{array}{cccc}
\hline
\multicolumn{4}{c}{\mathrm{Protein}}\\
\hline
\text{\textbf{ID}} & \text{Name} &  \text{\textbf{MW}} & \cdots\\
\hline
\mathbf{p_1} & \text{thioredoxin} &	\mathbf{11.8}& \cdots\\
\mathbf{p_2} & \text{flavodoxin}&	19.7& \cdots\\
\mathbf{p_3} & \text{ferredoxin}&	12.3& \cdots\\
\mathbf{p_4} & \text{ArgR} &	\mathbf{-700}& \cdots\\
\mathbf{p_5} & \text{CheW} &	18.1& \cdots\\
\vdots & \vdots & \vdots & \vdots \\
\hline
\end{array}
\quad
\begin{array}{cc}
\hline
\multicolumn{2}{c}{\mathrm{EnzymaticReaction}}\\
\hline
\mathbf{PID} & \mathbf{RID} \\
\hline
\mathbf{p_1} & \mathbf{r_1} \\
\mathbf{p_2} & \mathbf{r_1} \\
\mathbf{p_1} & \mathbf{r_2} \\
\mathbf{p_4} & \mathbf{r_2} \\
\mathbf{p_5} & \mathbf{r_3} \\
\vdots & \vdots \\
\hline
\end{array}
\quad\begin{array}{ccc}
\hline
\multicolumn{3}{c}{\mathrm{Reaction}}\\
\hline
\mathbf{ID} & \mathrm{Name} & \cdots\\
\hline
\mathbf{r_1} & \text{t-p + ATP = t d + ADP}& \cdots\\
\mathbf{r_2} & \text{H$_2$O + an a p $\to$ p + a c} & \cdots\\
\mathbf{r_3} & \text{D-r-5-p = D-r-5-p}& \cdots\\
\mathbf{r_4} & \text{$\beta$-D-g-6-p = f-6-p}& \cdots\\
\mathbf{r_5} & \text{p 4$'$-p + ATP = d-CoA + d} & \cdots\\
\vdots & \vdots & \vdots\\
\hline
\end{array}
\]
(b)
\begin{verbatim}
SELECT   R.Name as Name, AVERAGE(P.MW) as AvgMW 
FROM     Protein P, EnzymaticReaction ER, Reaction R
WHERE    P.ID = ER.ProteinID, ER.ReactionID = R.ID
GROUP BY R.Name
\end{verbatim}
(c)
\[
\begin{array}{cc}
\hline
\mathrm{Name} & \mathit{AvgMW} \\
\hline
\text{t-p + ATP = t d + ADP} &15.75\\
\text{H$_2$O + an a p $\to$ p + a c} & \textit{-338.2}\\
\text{D-r-5-p = D-r-5-p}&18.1\\
\vdots & \vdots \\
\hline
\end{array}\]

\caption{Example (a) input, (b) query, and (b) output data;
  input field names and values relevant to the \textit{italicized} erroneous output
  field or value are highlighted in \textbf{bold}.}\labelFig{example}
\end{figure}

 To illustrate our approach, consider the input data shown
in \refFig{example}(a) and the SQL query in \refFig{example}(b) which
calculates the average molecular weights of proteins involved in each
reaction.  The result of this query is shown in \refFig{example}(c).
The intuitive meaning of the SQL query is to find all
combinations of rows from the three tables Protein, EnzymaticReaction,
and Reaction such that the conditions in the WHERE-clause hold, then
group the results by the Name field, while averaging the MW (molecular
weight) field values and returning them in the AvgMW field.

Since the MW field contains the molecular weight of a protein, it is
clearly an error for the italicized value in the result to be
negative.  To track down the source of the error, it would be helpful
to know which parts of the input contributed to, or were
relevant to, the erroneous part of the output.  
We can
formalize this intuition by saying that a part of the output
depends on a part of the input if a change to the input part
may result in a change to the output part.  This is analogous to the
notion of dependence underlying program slicing~\cite{weiser81icse}, a
debugging aid that identifies the parts of a program on which a
program output may depend.

In this example, the input field values that the erroneous output
AvgMW-value depends on are highlighted in bold.  The dependencies
include the two summed MW values and the ID fields which are compared
by the selection and grouping query.  These ID fields must be included
because a change to any one of them could result in a change to the
italicized output value---for example, changing the occurrence of
$p_4$ in table EnzymaticReaction would change the average molecular
weight in the second row.  On the other hand, the names of the
proteins and reactions are irrelevant to the output
AvgMW---no changes to these parts can have any effect on the
italicized value, and so we can safely ignore these parts when looking
for the source of the error.

This example is simplistic, but the ability to concisely explain which
parts of the input influence each part of the output becomes more
important if we consider a large query to a realistic database with
tens or hundreds of columns per table and thousands or millions of
rows.  Manually tracing the dependence information in such a setting
would be prohibitively labor-intensive. Moreover, dependence information is
useful for a variety of other applications, including estimating the
\emph{freshness} of data in a query result by aggregating timestamps
on the relevant inputs, or transferring \emph{quality}  annotations
provided by users from the outputs of a query back to the inputs.

For example, suppose that each part of the database is annotated with
a timestamp.  Given a query, the dependence information shown in
\refFig{example} can be used to estimate the last modification time of
the data relevant to each part of the output, by summarizing the set
of timestamps of parts of the input contributing to an output part.
Similarly, suppose that the system provides users with the ability to
provide quality feedback in the form of star ratings.  If a user
flags the negative AvgMW value as being of low quality, this feedback
can be propagated back to the underlying data according to the
dependence information shown in \refFig{example} and provided to the
database maintainers who may find it useful in finding and correcting
the error.  In many cases, the user who finds the error may also be
the database maintainer, but dependency information still seems useful
as a debugging (or data cleaning) tool even if one has direct
access to the data.

In this article, we argue that data dependence provides a solid semantic
foundation for a provenance technique that highlights parts of the
input on which each part of the output depend.  We work in the setting
of the \emph{nested relational calculus}
(NRC)~\cite{DBLP:journals/sigmod/BunemanLSTW94,bntw,wong96jcss}, a
core language for database queries that is closely related to
\emph{monad algebra}~\cite{wadler92mscs}.  The NRC provides all of the
expressiveness of popular query languages such as SQL, and includes
\emph{collection types} such as sets or multisets, equipped with
union, comprehension, difference and equality operations.  The NRC can
also be extended to handle SQL's grouping and aggregation operations,
and functions on basic types.  We consider annotation-propagating
semantics for such queries and define a property called
\emph{dependency-correctness}, which, intuitively, means that the
provenance annotations produced by a query reflect how the output of
the query may change if the input is changed.

There may be many possible dependency-correct annotation-propagating
queries corresponding to an ordinary query.  In general, it is
preferable to minimize the annotations on the result, since this
provides more precise dependency information.  Unfortunately, as we
shall show, minimal annotations are not computable.  Instead,
therefore, we develop dynamic and static techniques that produce
dependency-correct annotations that are not necessarily minimal.  We
have implemented these techniques and found that they yield reasonable
results on small-scale examples; the implementation was used to
generate the results shown in \refFig{example}.

%--------------------------------------------------------------------%
\subsection{Prior Work on Provenance}\labelSec{related}
%--------------------------------------------------------------------%

We first review the relevant previous work on provenance and contrast it with
our approach.  We provide a detailed comparison with prior work on
program slicing and information flow in \refSec{discussion}.

%--------------------------------------------------------------------%
\subsubsection{Provenance for database queries} \labelSec{related-query-prov}
%--------------------------------------------------------------------%

Provenance for database queries has been studied by a number of
researchers, beginning in the early
1990s~\cite{buneman01icdt,buneman02pods,buneman08tods,DBLP:journals/tods/CuiWW00,green07pods,DBLP:conf/vldb/WangM90,woodruff97icde}.
Recent research on annotations, uncertainty, and incomplete
information~\cite{benjelloun06vldb,bhagwat05vldbj,DBLP:conf/icde/GeertsKM06}
has also drawn on these approaches to provenance; in particular,
definitions of provenance have been used to justify
annotation-propagation behaviors in these systems.  We will focus on
the differences between our work and the most recent work; please see
\citeN{buneman08pods} and \citeN{cheney09ftdb} for more complete
discussion of research on provenance in databases.

Most prior work on provenance has focused on identifying information
that explains \emph{why} some data is present in the output of a query
(or view) or \emph{where} some data in the output was copied from in
the input.  However, satisfying semantic characterizations of these
intuitions have proven elusive; indeed, many of the proposed definitions
themselves have been unclear or ambiguous.  Many proposed forms of
provenance are sensitive to query rewriting, in that equivalent
database queries may have different provenance behavior.  This raises
a number of troubling issues, since database systems typically rewrite
queries modulo equivalence, so the provenance of a query may change as
a result of query optimization.  Also, in part because of the absence
of clear formal definitions and foundations, these approaches have
been difficult to generalize beyond monotone relational queries in a
principled way.

% In the \emph{Polygen model} introduced by
% \citeN{DBLP:conf/vldb/WangM90}, each field of a table carries
% annotations consisting of sets of \emph{original} and
% \emph{intermediate} source tags.  The semantics of relational queries
% are adapted to propagate the annotations in an intuitively appealing,
% but essentially ad hoc way.

% \citeN{woodruff97icde} subsequently studied provenance (which they
% termed ``lineage'') in a database setting.  Their work focused on
% providing efficient techniques for associating parts of the output
% with parts of the input, with an emphasis on scientific data
% processing scenarios.  However, the exact semantics of the database
% operations was left to the user.

% \citeN{DBLP:journals/tods/CuiWW00} considered a model in which each
% \emph{record} in a table resulting from a database query is associated
% with a set of records in the input, called its lineage.  Lineage is
% defined according to a simple semantic criterion for each individual
% relational operation, and then lifted to arbitrary queries by
% composition; this, however, does not preserve the semantic criterion.
% Thus, \citeN{DBLP:journals/tods/CuiWW00}'s definition of lineage does
% not provide an obvious semantic guarantee and is sensitive to query
% rewriting, in the sense that equivalent queries may have different
% lineage.  

In \emph{why- and where-provenance}, introduced by
\citeN{buneman01icdt}, provenance is studied in a deterministic tree
data model, in which each part of the database can be addressed by a
unique path.  \citeN{buneman01icdt} considered two forms of
provenance: \emph{why-provenance}, which consists of a set of
witnesses, or subtrees of the input that suffice to explain a part
of the output, and \emph{where-provenance}, which consists of a
\emph{single} part of the input from which a given part of the output
was copied.  Both forms of provenance are sensitive to query rewriting
in general, but \citeN{buneman01icdt} discussed normal forms for
queries that avoid this problem.

\citeN{green07pods} showed that relations with semiring-valued
annotations on rows generalize several variations of the relational
model, including set, bag, probabilistic, and incomplete information
models, and identified a relationship between free semiring-valued
relations and why-provenance.  \citeN{foster08pods} extended this
approach to handle NRC queries and an unordered variant of XML.  These
approaches also appear orthogonal to our approach, and in additional
only consider annotations at the level of elements of collections, not
individual fields or collections, and they do not handle negation or
aggregation.

\citeN{buneman08tods} introduced a model of where-provenance for the
nested relational calculus.  In their approach each part of the
database is tagged with an optional annotation, or \emph{color}; colors
are propagated to the output so as to indicate where parts of the
output have been copied from in the input.  They studied the
expressiveness of this model compared to queries that explicitly
manipulate annotations.  They also investigated where-provenance for
updates, which we discuss in \refSec{related-update-prov}.

Our work is closest in spirit to the why-provenance and lineage
techniques; however, in contrast to these techniques our approach
annotates every part of the database and provides clear semantic
guarantees and qualitatively useful provenance information in the
presence of negation, grouping and aggregation.

%--------------------------------------------------------------------%
\subsubsection{Provenance for database updates} \labelSec{related-update-prov}
%--------------------------------------------------------------------%
Some recent work has generalized where-provenance to database
updates~\cite{DBLP:conf/sigmod/BunemanCC06,buneman08tods}, motivated
by \emph{curated} scientific databases that are updated frequently, often
by (manual) copying from other sources.  This work has focused on
recording the external sources of the data in a database
and tracking how the data has been rearranged within a database across
successive versions.  Accordingly, the provenance information provided
by these approaches only connects data to exact copies in other
locations, and does not track provenance through other operations.  In
this sense, it is similar to the where-provenance approach
considered by~\citeN{buneman01icdt} for database queries.

Our approach addresses an orthogonal issue, that of understanding how
data in the result of a query depends on parts of the input; we
therefore track provenance through copies as well as other forms of
computation.  Although our definition of dependency correctness could
also be used for updates, it is not clear whether this yields a useful
form of provenance, and we plan to investigate alternative dependency
conditions that are more suitable for updates, using the update
language employed in~\cite{buneman08tods}.

%--------------------------------------------------------------------%
\subsubsection{Workflow provenance} 
%--------------------------------------------------------------------%
Provenance has also been studied in geospatial and scientific
computation~\cite{bose05cs,ipaw06,DBLP:journals/sigmod/SimmhanPG05},
particularly for \emph{workflows} (visual programs written by
scientists).  In their simplest form (see e.g. the Provenance
Challenge~\cite{Editorial:Challenge06}), workflows are essentially
directed acyclic graphs (DAGs) representing a computation.  For such DAG
workflows, the provenance information that is typically stored is
simply the workflow DAG, annotated with additional information, such
as filenames and timestamps, describing the arguments that were used
to compute the result of interest.  This corresponds to a simple form
of dependency tracking, although as far as we know no research on
workflow provenance has explicitly drawn this connection.

However, many more sophisticated workflow programming models have been
developed, involving concurrency and distributed computation.  For
these models, the appropriate correctness criteria for provenance
tracking are much less clear.  In fact, the ordinary semantics of
these systems is not always clearly specified.  One principled
approach recently introduced by \citeN{DBLP:conf/dils/HiddersKSTB07}
defines provenance for the nested relational calculus augmented with
additional function symbols that represent calls to scientific
workflow components.  Their approach has so far focused on
defining provenance and not formulating or proving desirable correctness
properties.  We believe dependence analysis may provide
an appropriate foundation for provenance in this and other workflow
programming models.

\subsection{Contributions}

%\todo{Explain contributions more carefully.}

The main contribution of this article is the development of a semantic
criterion called \emph{dependency-correctness} that characterizes a
form of provenance information we call \emph{dependency provenance}.
Dependency-correctness captures an intuition that provenance should
link a part of the output to all parts of the input on which the
output part depends, enabling us to make some predictions about the
effects of changes to the input and to quickly identify source data
that contributed to an error in the output.

Building on this framework, we show that (unsurprisingly) it is
undecidable whether some dependency-correct provenance information is
minimal, and proceed to develop computable dynamic and static
techniques for conservatively approximating correct dependency
provenance.  These techniques, and their correctness proofs, are
largely standard but the presence of database query language features
and collection types introduces complications that have not been
addressed before in work on information flow or program slicing.

\subsection{Organization}
The structure of the rest of this article is as follows.  We review the
syntax, type system and semantics of the nested relational calculus in
\refSec{background}.  We then introduce (in
\refSec{provenance-dependence}) the annotation-propagation model,
motivate and define dependency-correctness, and show that it is
impossible to compute minimal dependency-correct annotations.  In
\refSec{provenance-tracking} we describe a dynamic
\emph{prov\-en\-ance-tracking} semantics that is dependency-correct.
We also (\refSec{provenance-analysis}) introduce a static, type-based
\emph{provenance analysis} which is less accurate than provenance
tracking, but can be performed statically; we also prove its
correctness relative to dynamic provenance tracking.  We discuss
experience with a prototype implementation in \refSec{discussion} and
discuss future work and conclude in \refSec{concl}.

%--------------------------------------------------------------------%
\section{Background}\labelSec{background}
%--------------------------------------------------------------------%

We will provide a brief review of the \emph{nested relational
  calculus} (NRC)~\cite{bntw}, a core database query language which is
closely related to \emph{monad algebra}~\cite{wadler92mscs}.  The
nested relational calculus is a typed functional language with types
$\tau$ of the form:
\begin{eqnarray*}
  \tau &::=& \boolTy \mid \intTy  \mid \tau_1 \times \tau_2  \mid \setof{\tau}
\end{eqnarray*}
We consider base types $\boolTy$ and $\intTy$, along with product
types $\tau_1 \times \tau_2$ and \emph{collection types}
$\setof{\tau}$.  Collection types typically are taken to be monads
equipped with an addition operator (sometimes called \emph{ringads});
typical examples used in databases include lists, sets, or multisets,
and in this article we consider finite multisets (also known as bags).

The expressions of our variant of NRC are as follows:
\begin{eqnarray*}
  e &::=& x \mid \letin{x=e_1}{e_2} \mid (e_1,e_2) \mid \pi_1(e) \mid \pi_2(e) \\
  &\mid& b  \mid \neg e \mid e_1 \andd e_2
   \mid  e_1 \eq e_2\mid\ifthenelse{e_0}{e_1}{e_2}  \\
&\mid& i \mid e_1 + e_2 \mid \ksum(e)  \\
  &\mid& \emptyset \mid \setof{e}\mid e_1 \cup e_2 \mid e_1 - e_2\mid \{e_2 \mid x \in e_1\} \mid \bigcup e
\end{eqnarray*}
Here, $i \in \Int = \{\ldots,-1,0,1,\ldots\}$ denotes integer
constants and $b \in \Bool = \{\ktrue,\kfalse\}$ denotes Boolean
constants.  The bag operations include $\emptyset$, the constant empty
multiset; singletons $\{e\}$; multiset union $\cup$, difference $-$,
and comprehension $\{e_2 \mid x \in e_1\}$; and flattening $\bigcup
e$.  By convention, we write $\{e_1,\ldots,e_n\}$ as syntactic sugar
for $\{e_1\} \cup \cdots \cup \{e_n\}$.  Finally, we include $\ksum$,
a typical \emph{aggregation} operation, which adds together all of the
elements of a multiset and produces a value; e.g.  $\ksum \{1,2,3\} =
6$.  By convention, we take $\ksum (\emptyset) = 0$.  We syntactically
distinguish between NRC's equality operation $\eq$ and mathematical
equality $=$.

\begin{figure}[p]
\fbox{$\wf{\Gamma}{e}{\tau}$}
\[\begin{array}{c}
\infer{\wf{\Gamma}{x}{\tau}}{x{:}\tau \in \Gamma}
\quad 
\infer{\wf{\Gamma}{\letin{x=e_1}{e_2}}{\tau_2}}
{\wf{\Gamma}{e_1}{\tau_1} & \wf{\Gamma,x{:}\tau_1}{e_2}{\tau_2}}
\quad
\infer{\wf{\Gamma}{i}{\kint}}{i \in \Int}
\quad
\infer{\wf{\Gamma}{e_1 + e_2}{\intTy}}{\wf{\Gamma}{e_1}{\intTy}&\wf{\Gamma}{ e_2}{\intTy}}
\smallskip\\
\infer{\wf{\Gamma}{\ksum(e)}{\intTy}}{\wf{\Gamma}{e}{\setof{\intTy}}}
\quad
\infer{\wf{\Gamma}{b}{\kbool}}{b \in \Bool}
\quad
\infer{\wf{\Gamma}{\ifthenelse{e_0}{e_1}{e_2}}{\tau}}
{\wf{\Gamma}{e_0}{\kbool} 
& \wf{\Gamma}{e_1}{\tau}
& \wf{\Gamma}{e_2}{\tau}}
\quad
\infer{\wf{\Gamma}{\neg e}{\boolTy}}{\wf{\Gamma}{e}{\boolTy}}
\smallskip\\
\infer{\wf{\Gamma}{e_1 \andd e_2}{\boolTy}}{\wf{\Gamma}{e_1}{\boolTy}&\wf{\Gamma}{ e_2}{\boolTy}}
\quad
\infer{\wf{\Gamma}{(e_1,e_2)}{\tau_1 \times \tau_2}}
{
\wf{\Gamma}{e_1}{\tau_1}
&
\wf{\Gamma}{e_2}{\tau_2}
}
\quad
\infer[(i \in\{1,2\})]{\wf{\Gamma}{\pi_i(e)}{\tau_i}}{\wf{\Gamma}{e}{\tau_1 \times \tau_2}}
\quad
\smallskip\\
\infer{\wf{\Gamma}{e_1 \eq e_2}{\boolTy}}{\wf{\Gamma}{e_1}{\tau}&\wf{\Gamma}{ e_2}{\tau} }
\quad
\infer{\wf{\Gamma}{\emptyset}{\setof{\tau}}}{}
\quad 
\infer{\wf{\Gamma}{\setof{e}}{\setof{\tau}}}{\wf{\Gamma}{e}{\tau}}
\quad
\infer{\wf{\Gamma}{e_1 \cup e_2}{\setof{\tau}}}
{\wf{\Gamma}{e_1 }{\setof{\tau}}& \wf{\Gamma}{e_2 }{\setof{\tau}}}
\smallskip\\
\infer{\wf{\Gamma}{e_1 - e_2}{\setof{\tau}}}
{\wf{\Gamma}{e_1 }{\setof{\tau}}& \wf{\Gamma}{e_2 }{\setof{\tau}}}
\quad
\infer{\wf{\Gamma}{\{e_2 \mid x \in e_1\}}{\setof{\tau_2}}}
{\wf{\Gamma}{e_1}{\setof{\tau_1}} & \wf{\Gamma,x{:}\tau_1}{e_2}{\tau_2}}
\quad 
\infer{\wf{\Gamma}{\bigcup{e}}{\setof{\tau}}}
{ \wf{\Gamma}{e}{\setof{\setof{\tau}}}}
\end{array}
\]
\caption{Well-formed query expressions}\labelFig{expression-typing}
\[\begin{array}{rclcrcl}
\E{x}\gamma &=& \gamma(x)&&
\E{\letin{x=e_1}{e_2}}\gamma&=& \E{e_2}\gamma[x \mapsto \E{e_1}\gamma]\\
\E{i}\gamma &=& i&&
\E{e_1 + e_2}\gamma &=& \E{e_1}\gamma + \E{e_2}\gamma\\
\E{\ksum(e)}\gamma &=& \sum \E{e}\gamma&&
\E{b}\gamma &=& b\\
\E{\nott e}\gamma &=& \nott\E{e}\gamma&&
\E{e_1 \andd e_2}\gamma &=& \E{e_1}\gamma \andd \E{e_2}\gamma\\
\E{(e_1,e_2)}\gamma &=& (\E{e_1}\gamma ,\E{e_2}\gamma )&&
\E{\pi_i(e)}\gamma &=& \pi_i(\E{e}\gamma) \quad (i \in \{1,2\})\\
\E{\emptyset}\gamma &=& \emptyset&&
\E{\setof{e}}\gamma &=& \{\E{e}\gamma\}\\
\E{e_1 \cup e_2}\gamma &=& \E{e_1}\gamma \cup \E{e_2}\gamma&&
\E{e_1 - e_2}\gamma &=& \E{e_1}\gamma - \E{e_2}\gamma\\
\E{\bigcup e}\gamma &=& \bigcup{\E{e}\gamma}&&
\E{\{e \mid x \in e_0\}}\gamma &=& \{\E{e}\gamma[x\mapsto v] \mid v \in \E{e_0}\gamma\}
\end{array}\]
\begin{eqnarray*}
\E{\ifthenelse{e_0}{e_1}{e_2}}\gamma = \left\{ 
\begin{array}{ll}
\E{e_1}\gamma & \text{if $\E{e_0}{\gamma} = \ktrue$}\\
\E{e_2}\gamma & \text{if $\E{e_0}{\gamma} = \kfalse$}
\end{array}\right.\\
\E{e_1 \eq e_2}\gamma = \left\{ 
\begin{array}{ll}
\ktrue & \text{if $\E{e_1}{\gamma} = \E{e_2}{\gamma}$}\\
\kfalse& \text{if $\E{e_1}{\gamma} \neq \E{e_2}{\gamma}$}
\end{array}\right.
\end{eqnarray*}
\caption{Semantics of query expressions}\labelFig{semantics}
\begin{eqnarray*}
\Pi_A(R) &=& \{x.A \mid x \in R\}\\
\sigma_{A=B}(R) &=& \bigcup\{\ifthenelse{x.A=x.B}{\{x\}}{\emptyset} \mid x \in R\}\\
R \times S &= & \{(A:x.A,B:x.B,C:y.C,D:y.D,E:y.E) \mid x \in R, y \in S\}\\
\Pi_{BE}(\sigma_{A=D}(R\times S)) &= & \{\ifthenelse{x.A=y.D}{\{(B:x.B,E:y.E)\}}{\emptyset} \mid x \in R, y \in S\}\\
R\cup \rho_{A/C,B/D}(\Pi_{CD}(S)) &=& R \cup \{(A:y.C,B:y.D)\mid y \in S\}\\
R- \rho_{A/D,B/E}(\Pi_{DE}(S)) &=& R - \{(A:y.D,B:y.E)\mid y \in S\}\\
\ksum(\Pi_A(R)) &=& \ksum\{x.A \mid x \in R\}\\
\kcount(R) &=& \ksum\{1 \mid x \in R\}
\end{eqnarray*}
\caption{Example queries}\labelFig{example-queries}
\end{figure}

NRC expressions can be typechecked using standard
techniques.  Contexts $\Gamma$ are lists of pairs of variables
and types $x_1:\tau_1,\ldots,x_n:\tau_n$, where $x_1,\ldots,x_n$ are distinct.
The rules for typechecking expressions are shown in
\refFig{expression-typing}.  

We write $\mathcal{M}_{\kfin}(X)$ for the set of all finite multisets
with elements drawn from $X$.  The (standard) interpretation of base
types as sets of values is as follows:
\[\begin{array}{rcl}
\T{\kbool} &=& \Bool = \{\ktrue,\kfalse\}
\smallskip\\
\T{\kint} &=& \Int = \{\ldots,-1,0,1,\ldots\}
\smallskip\\
\T{\tau_1\times\tau_2} &=& \T{\tau_1}\times \T{\tau_2}
\smallskip\\
\T{\setof{\tau}} &=& \mathcal{M}_{\kfin}(\T{\tau})
\end{array}\]
An environment $\gamma$ is a function from variables to values.  We
define the set of environments matching context $\Gamma$ as
$\T{\Gamma} = \{\gamma \mid \forall x \in \dom(\Gamma).~\gamma(x) \in 
\T{\Gamma(x)}\}$.

\refFig{semantics} gives the semantics of queries.  Note that we
overload notation for pair projection $\pi_i$ and bag operations such
as $\cup$ and $\bigcup$; also, if $S$ is a bag of integers, then $\sum
S$ is the sum of their values (taking $\sum\emptyset = 0$).  It is
straightforward to show that
\begin{lemma}
  If $\wf{\Gamma}{e}{\tau}$ then $\E{e} : \T{\Gamma} \to \T{\tau}$.
\end{lemma}

\begin{remark}
  As discussed in previous work~\cite{bntw}, the NRC can express a
  wide variety of queries including ordinary relational queries,
  nested subqueries, and grouping and aggregation queries.  The core
  NRC excludes a number of convenient features such as records with
  named fields and comprehensions.  However, these features can be
  viewed as syntactic sugar for core NRC expressions.  In particular,
  we use abbreviations such as:
\begin{eqnarray*}
  \{e \mid x_1 \in e_1, x_2 \in e_2\} &=& 
  \bigcup\{\{e \mid x_2 \in e_2\} \mid x_1 \in e_1\}\\
  \{e \mid x \in e_0, C\} &=& 
  \bigcup\{\ifthenelse{C}{\{e\}}{\emptyset} \mid x \in e_0\} \\
  \{e \mid (x_1,x_2) \in e_0\} &=& 
  \{\letin{x_1=\pi_1(x), x_2=\pi_2(x)}{e} \mid x \in e_0\}
\end{eqnarray*}

Additional base types, primitive functions and relations such as
$\kreal$, $/ : \kreal \times \kreal \to \kreal$, and $\kaverage :
\{\kreal\} \to \kreal$ can also be added without difficulty.  For
example, using more readable named records, comprehensions, and
pattern-matching, the SQL query from\refFig{example}(b) can be
defined as
\[\begin{array}{l}
\letin{X = \{(r.Name,p.MW) \mid 
	r \in R, 
	er \in ER, 
	p \in P, 
	er.RID = r.ID,
	p.ID = er.PID\}}{\\
\{(n, \kaverage \{mw\mid (n',mw) \in X, n=n'\}) \mid (n,\_) \in X\}}\end{array}\]
Additional examples are
shown in \refFig{example-queries}.  

We do not consider other features of SQL such as operator overloading
or incomplete information (\verb|NULL| values).
\end{remark}

%--------------------------------------------------------------------%
\section{Annotations, Provenance and
  Dependence}\labelSec{provenance-dependence}
%--------------------------------------------------------------------%

We wish to define \emph{dependency provenance} as information relating
each part of the output of a query to a set of parts of the input on
which the output part depends.  Collection types such as sets and bags
are unordered and lack a natural way to address parts of values, so we
must introduce one.  One technique (familiar from many program
analyses~\cite{nielson05ppa} as well as other work on
provenance~\cite{buneman08tods,DBLP:conf/vldb/WangM90}) is to enrich
the data model with \emph{annotations} that can be used to refer to
parts of the value.  In practice, the annotations might consist of
explicit paths or addresses pointing into a particular representation
of a part of the data, analogous to filenames and line number
references in compiler error messages, but for our purposes, it is
preferable to leave the structure of annotations abstract; we
therefore consider annotations to be sets of \emph{colors}, or
elements of some abstract data type $\Color$.

We can then infer provenance information from functions on annotated
values by observing how such functions propagate annotations;
conversely, we can define provenance-tracking semantics by enriching
ordinary functions with annotation-propagation behavior.  However, for
any ordinary function, there are many corresponding
annotation-propagating functions so the question arises of how to
choose among them.  

We consider two natural constraints on the annotated functions we will
consider.  First, if we ignore annotations, the behavior of an
annotated function should correspond to that of an ordinary function.
Second, the behavior of the annotated functions should treat the
annotations abstractly, so that we may view the colors as locations.
We show that both properties follow from a single condition
called \emph{color-invariance}.

We next define dependency-correctness, a property characterizing
annotated functions whose annotations safely over-approximate the
dependency behavior of some ordinary function.  Such annotations can
be used to compute a natural notion of ``data slices'', by
highlighting those parts of the input on which a given part of the
output \emph{may} depend.  It is clearly desirable to produce slices
that are as small as possible.  Unfortunately, minimal slices turn out
not to be computable since it is undecidable whether the annotations
in the output of a dependency-correct function are minimal.  In the
next sections, we will show how to calculate approximate dynamic
and static dependency information for NRC queries.

We define \emph{annotated values (a-values)} $v$, \emph{raw values
  (r-values)} $w$, and \emph{multisets of annotated values} $V$ as
follows:
\begin{eqnarray*}
v &::=& w^\Phi \qquad w ::= i \mid b \mid (v_1,v_2) \mid V \qquad V ::= \{v_1,\ldots,v_n\}
\end{eqnarray*}
Annotations are \emph{sets} $\Phi \subseteq \Color$ of values from
some atomic data type $\Color$, called \emph{colors}.  We often omit
set brackets in the annotations, for example writing $w^{a,b,c}$
instead of $w^{\{a,b,c\}}$ and $w$ instead of $w^\emptyset$.  An
a-value $v$ is said to be \emph{distinctly colored} if every part of
it is colored with a singleton set $\{a\}$ and no color $c$ is used
more than once in $v$.

For each type $\tau$, we define the set $\AAz{\tau}$ of annotated values of type $\tau$ as follows:
\[
\begin{array}{rcl}
  \AAz{\boolTy} &=& \{b^\Phi \mid b \in \Bool\}\\
  \AAz{\intTy} &=& \{i^\Phi \mid i \in \Int\}\\
  \AAz{\tau_1\times\tau_2} &=& \{(v_1,v_2)^\Phi \mid v_1 \in \AAz{\tau_1},v_2 \in \AAz{\tau_2}\}\\
  \AAz{\setof{\tau}} &=& \{V^\Phi \mid \forall v \in V. v \in \AAz{\tau}\}
\end{array}
\]
Annotated environments $\hg$ map variables to annotated values.  We
define the set of annotated environments matching context $\Gamma$ as
$\AAz{\Gamma} = \{\hg \mid \forall x \in \dom(\Gamma). \hg(x) \in
\AAz{\Gamma(x)}\}$.

We define an \emph{erasure function} $|-|$, mapping a-values to
ordinary values (and, abusing notation, also mapping r-values to
ordinary values), and an \emph{annotation extraction function} $\|-\|$
which extracts the set of all colors mentioned anywhere in an a-value
or r-value, as follows:
\[
\begin{array}{rcl}
|i| &=& i 
\\
|b| &=&b 
\\
|(v_1 ,v_2)| &=& (|v_1|, |v_2|) 
\\
|\{V\}| &=& \{|v| \mid v \in V\}
\\
|w^\Phi| &=&  |w|
\end{array}
\qquad
\begin{array}{rcl}
\|i\| &=& \emptyset
\\ 
\|b\| &=&\emptyset 
\\ 
\|(v_1 , v_2)\| &=& \|v_1\| \cup \|v_2\| 
\\ 
\|\{V\}\| &=& \bigcup\{\|v\| \mid v \in V\}
\\
\|w^\Phi\| &=& \Phi \cup \|w\|
\end{array}
\]
Two a-values are said to be \emph{compatible} (written $v \cong v'$) if
$|v| = |v'|$; also, an a-value $v$ is said to \emph{enrich} an
ordinary value $v'$ (written $v \enriches v'$) provided $|v| = v'$.

We now consider annotated functions (a-functions) $F : \AAz{\tau} \to
\AAz{\tau'}$ on a-values.  We say that a-function $F$ enriches an
ordinary function $f: \T{\tau} \to \T{\tau'}$ (written $F \enriches
f$), provided that $\forall v\in\AAz{\tau}.f(|v|) = |F(v)|$.  We will
also consider annotated functions $F : \AAz{\Gamma} \to \AAz{\tau}$
mapping annotated environments to values.  We say that an a-function $F$
enriches an ordinary function $f: \T{\Gamma} \to \T{\tau}$ (again
written $F \enriches f$), provided that $\forall
\gamma\in\AAz{\Gamma}.f(|\gamma|) = |F(\gamma)|$.

%--------------------------------------------------------------------%
\subsection{Color-invariance} 
%--------------------------------------------------------------------%

Clearly, many exotic a-functions exist that are not enrichments of
any ordinary function.  For example, consider
\[
F(i^\Phi) = \left\{\begin{array}{ll}
1^\Phi & (\Phi = \emptyset)\\
0^\Phi & (\Phi \neq \emptyset)
\end{array}\right.
\]
Here, $F$ tests whether its annotation is empty or not, and there is
no ordinary function $f : \T{\intTy} \to \T{\intTy}$ such that $\forall v.
|F(v)| = f(|v|)$.  In the rest of this article we will restrict
attention to a-functions $F$ that are enrichments of ordinary
functions.

In fact, we will restrict attention still further to a-functions whose
behavior on colors is also abstract enough to be consistent with an
interpretation of colors in the input as addresses for parts of the
input.  For example, consider $G,H : \AAz{\intTy} \to \AAz{\intTy}$
having the following behavior:
\begin{eqnarray*}
  G(i^\Phi) &= &\left\{\begin{array}{ll}
      i^\emptyset & (\Phi = \emptyset)\\
      i^{\{c\}} & (\Phi \neq \emptyset)
\end{array}\right.\\
  H(i^\Phi) &=& i^{\Phi - \{c\}}
\end{eqnarray*}
where in both cases $c$ is some fixed color.  Both functions are
enrichments of the ordinary identity function on integers, $\lambda
i.i$, but both perform nontrivial computations on the annotations. If
we wish to interpret the colors on these functions as representing
sets of locations, then we want to exclude from consideration
functions like $G$ whose behavior depends on the size of the
annotation set or functions like $H$ whose behavior depends on a
specific color.

By analogy with generic queries in relational databases~\cite{ahv},
such a-functions ought to behave in a way that is insensitive to the
particular choice of colors.  Moreover, since a-values are annotated
by sets of colors, the a-functions also ought to be insensitive to
properties of the annotations such as nonemptiness or equality.  In
particular, we expect that the behavior of an a-function is
determined by its behavior on distinctly-colored inputs.

To make this precise, we first need to define some auxiliary concepts.
\begin{definition}
  An a-value $v$ is \emph{distinctly-colored} provided every
  subexpression $w^{\Phi}$ we have $\Phi = \{c\}$ for some color $c$,
  and no two subexpressions occurring in $v$ have the same color.
\end{definition}

\begin{example}
  For example, $v = \{(1^a,1^b)^c\}^d$ is distinctly-colored, while
  $v' = \{(1^a,1^a)^c\}^d$ is not, because the color $a$ is re-used in
  two different subexpression occurrences of $1^a$.
\end{example}

A \emph{color substitution} is a function $\alpha : \colorTy \to
\{\colorTy\}$ mapping colors to sets of colors.  We can lift color
substitutions to act on arbitrary a-values as follows:
\begin{eqnarray*}
  \alpha(b) &=& b\\
  \alpha(i) &=& i\\
  \alpha(v_1,v_2) &=& (\alpha(v_1),\alpha(v_2))\\
  \alpha(V) &=& \{\alpha(v) \mid v \in V\}\\
\alpha(w^\Phi) &=& (\alpha(w))^{\alpha[\Phi]}
\end{eqnarray*}
where $\alpha[\Phi] = \bigcup\{\alpha(c) \mid c \in \Phi\}$.  Note
that for any $v \in \AAz{\tau}$, we have $\alpha(v) \in \AAz{\tau}$;
we sometimes write $\alpha^\tau$ to indicate the restriction of
$\alpha$ to $\AAz{\tau}$.

\begin{example}
  Continuing the previous example, consider the color substitution
  defined by $\alpha(a) = \{b,c\}$ and $\alpha(x) = \{x\}$ for $x \neq
  a$.  Applying this substitution to $v$ yields
  $\{(1^{a,b},1^b)^c\}^d$.  Applying to $\{1^{a}, 2^{a,d}\}^c$ yields
  $\{1^{b,c}, 2^{b,c,d}\}^c$.
\end{example}

We note some useful properties relating distinctly-colored values,
color-substitution and the erasure and color-support functions; these
are easy to prove by induction.
\begin{lemma}
Suppose $\alpha : \colorTy \to \{\colorTy\}$. Then (1) $|\alpha(v)| =
|v|$ and (2) $\|\alpha(v)\| = \alpha[\|v\|]$.
\end{lemma}
\begin{lemma}
  Suppose $v$ is an a-value.  Then there exists a distinctly-colored
  $v_0 \cong v$ and a color substitution $\alpha_0$ such that
  $\alpha_0(v_0) = v$.
\end{lemma}
Accordingly, for each ordinary value $v$ fix a distinctly-annotated
$\dc(v)$; moreover, for each a-value $v$ fix a color substitution
$\alpha_v$ such that $\alpha_v(\dc|v|) = v$.

We now define a property called \emph{color-invariance}, by analogy
with the \emph{color-propagation} studied in~\cite{buneman08tods} for
annotations consisting of single colors.  Color-invariance is defined
as follows.
\begin{definition}[Color-invariance]
  An a-function $F:\AAz{\tau_1} \to \AAz{\tau_2}$ is called
  \emph{color-invariant} if whenever $\alpha : \colorTy \to
  \{\colorTy\}$ then we have $\alpha^{\tau_2} (F(v)) = F(\alpha^{\tau_1}(v))$.
\end{definition}

As noted above, color-invariance has two important consequences.
First, the behavior of a color-invariant function is determined by its
behavior on distinctly-colored inputs.  Second, color-invariant
functions are always enrichments of ordinary functions.

\begin{proposition}
  If $F,G : \AAz{\tau_1} \to \AAz{\tau_2}$ is color-invariant then the following are equivalent:
\begin{enumerate}
\item $F = G$
\item  $F(v) = G(v)$ for every
  \emph{distinctly-colored} $v\in \AAz{\tau_1}$.  
\item  $F(\dc(v)) = G(\dc(v))$ for every ordinary value $v \in
  \T{\tau_1}$.
\end{enumerate}
\end{proposition}
\begin{proof*}
  The implications $(1) \Rightarrow (2) \Rightarrow (3)$ are trivial.  We show (3)
  implies (1).  Let $v\in \AAz{\tau_1}$ be given.  Then $v =
  \alpha_v(\dc(|v|))$, so to prove $F = G$, it suffices to show:
  \[F(v) = F(\alpha_v(\dc(|v|))) = \alpha_v (F(\dc(|v|))) =
  \alpha_v(G(\dc(|v|))) = G(\alpha_v(\dc(|v|))) =  G(v)\]
\end{proof*}
\begin{proposition}
  If $F : \AAz{\tau_1} \to \AAz{\tau_2}$ is color-invariant then $F
  \enriches f$ where $f(v) = |F(\dc(v))|$.
\end{proposition}
\begin{proof*}
  Let $v \in \AAz{\tau_1}$ be given.  Then to prove $F \enriches f$, observe:
  \[
  f (|v|) = |F(\dc(|v|))| = |\alpha_v(F(\dc(|v|)))| = |F(\alpha_v(\dc(|v|)))|
  = |F(v)|
  \]

\end{proof*}

We write $|F|$ for $f$ provided $F \enriches f$; clearly, $f$ is
unique when it exists, and $|F|$ exists for any color-invariant $F$.

%--------------------------------------------------------------------%
\subsection{Dependency-correctness} 
%--------------------------------------------------------------------%

We now turn to the problem of characterizing a-functions whose
annotation behavior captures a form of dependency information.
Intuitively, an a-function $F$ is dependency-correct if its output
annotations tell us how changes to parts of the input may affect parts
of the output.  First, we need to capture the intuitive notion of
changing a specific part of a value.  
\begin{definition}[Equal except at $c$]
  Two a-values $v_1,v_2$ are \emph{equal except at $c$} ($v_1 \equiv_c v_2$)
  provided that they have the same structure except possibly at subterms
  labeled with the color $c$; this relation is defined as follows:
\[\begin{array}{c}
\infer{d \equiv_c d}{d \in \Bool \cup \Int}
\quad %\smallskip\\
\infer{(v_1,v_2) \equiv_c (v_1',v_2')}{v_1 \equiv_c v_1' & v_2 \equiv_c v_2'}
\quad %\smallskip\\
\infer{\{v_1,\ldots,v_n\} \equiv_c \{v_1',\ldots,v_n'\}}{v_1 \equiv_c v_1' & \cdots & v_n  \equiv_c v_n'}
\quad\infer{w_1^{\Phi} \equiv_c w_2 ^{\Phi}}{w_1 \equiv_c w_2}
\quad %\smallskip\\
\infer{w_1^{\Phi_1} \equiv_c w_2 ^{\Phi_2}}{c \in \Phi_1 \cap \Phi_2}
\end{array}
\]
Furthermore, we say that two annotated environments $\hg,\hg'$ are
equal except at $a$ (written $\hg \equiv_a \hg'$) if their domains are
compatible ($\dom(\hg) = \dom(\hg')$) and they are pointwise equal
except at $a$, that is, for each $x \in \dom(\hg)$, we have $\hg(x)
\equiv_a \hg'(x)$.
\end{definition}
\begin{remark}
  For distinctly-colored values, a color serves as an address uniquely
  identifying a subterm.  Thus, $\equiv_c$ relates a
  distinctly-colored value to a value which can be obtained by
  modifying the subterm located at $c$; that is, if we write $v_1$ as
  $C[v_1']$ where $C$ is a context and $v_1'$ is the subterm labeled
  with $c$ in $v_1$, and $v_1 \equiv_c v_2$, then $v_2 = C[v_2']$ for
  some subterm $v_2'$ labeled with $c$.  Note that $v_2'$ and $v_2$
  need not be distinctly colored, and that $\equiv_c$ makes sense for
  arbitrary a-values, not just distinctly colored ones.
\end{remark}
\begin{example}
  Consider the two a-environments:
  \begin{eqnarray*}
    \hg&=&(\mathrm{R}:\{(\mathrm{1}^{c_1},\mathrm{3}^{c_2},5^{c_3})^{b_1},\ldots\}^a, \mathrm{S}:\cdots)
    \\
    \hg'&=&(\mathrm{R}:\{(\mathrm{2}^{c_1},\mathrm{3}^{c_2},5^{c_3})^{b_1},\ldots\}^a,\mathrm{S}:\cdots)
\end{eqnarray*}
We have $\hg \equiv_a \hg'$, $\hg \equiv_{b_1} \hg'$, and $\hg
\equiv_{c_1} \hg'$, assuming that the elided portions are identical.
\end{example}

\begin{definition}[Dependency-correctness]
  An a-function $F:\AAz{\Gamma}\to\AAz{\tau}$ is
  \emph{de\-pend\-en\-cy-correct} if for any $c \in \Color$ and $\hg,\hg'
  \in \AAz{\Gamma}$ satisfying $\hg \equiv_c \hg'$, we have $F(\hg)
  \equiv_c F(\hg')$.
\end{definition}

\begin{example}
  Recall $\hg,\hg'$ as in the previous example.  Suppose $F$ is
  dependency-correct and
  \[F(\hg) =
  \{(1^{c_1},3^{c_2},5^{c_3})^{b_1}\}^a\;.\]
  Since $\hg \equiv_{c_1} \hg'$, we know that $
  F(\hg)\equiv_{c_1}F(\hg')$
  so we can see that $F(\hg')$ must be of the form
  \[\{(x^{c_1},3^{c_2},5^{c_3})^{b_1}\}^a \] 
  for some $x \in \Int$.  We do \emph{not} necessarily know that $x$
  must be $2$; this is not captured by
  dependency-correctness.
\end{example}

\begin{remark}
  Dependency-correctness tells us that for any $c$, we must have
  $F(\hg) = C[v_1,\ldots,v_n]$ and $F(\hg') = C[v_1',\ldots,v_n']$,
  where $C[-,\ldots,-]$ is a context not mentioning $c$ and
  $v_1,\ldots,v_n$, $v_1',\ldots,v_n'$ are labeled with $c$.  Thus, $F$'s
  annotations tell us which parts of the output (i.e.,
  $v_1,\ldots,v_n$) \emph{may} change if the input is changed at $c$.
  Dually, they also tell us what part of the output (i.e.,
  $C[-,\ldots,-]$) \emph{cannot} be changed by changing the input at
  $c$.  

  We can consider the parts of the output labeled with $c$ to be a
  \emph{forward slice} of the input at $c$; it shows all of the parts
  of the output that may depend on $c$.  Conversely, suppose the
  output is of the form $C'[w^\Phi]$.  Then we can define a
  \emph{backward slice} corresponding to this part of the output by
  factoring $\hg$ into $C[v_1,\ldots,v_n]$ where $C$ is as small as
  possible subject to the constraint that $\Phi \cap \|v_i\| =
  \emptyset$ for each $i$.  This context $C[-,\ldots,-]$ identifies
  all of the parts of the input on which a given part of the output
  may depend.
\end{remark}

Of course, dependency-correctness does not uniquely characterize the
annotation behavior of a given $F$.  It is possible for the
annotations to be dependency-correct but inaccurate.  For
example we can always trivially annotate each part of the output with
every color appearing in the input.  This, of course, tells us
nothing about the function's behavior.  In general, the fewer the
annotations present in the output of a dependency-correct $F$, the
more they tell us about $F$'s behavior.  We therefore consider a
function $F$ to be \emph{minimally annotated} if no annotations can be
removed from $F$'s output for any $v$ without damaging correctness.

\begin{example}
  For example, consider the ordinary function 
\[f(x,y) = \left\{\begin{array}{ll}
    y &: x = 0\\
    x+1 &: x \neq 0
  \end{array}
\right.\;.
\]
Then the function 
\[F(x^a,y^b) = \left\{\begin{array}{ll}
    y^{a,b} &: x = 0\\
    (x+1)^{a,b} &: x \neq 0
  \end{array}
\right.\;.
\]
%$F(x^a,y^b) = \ifthenelse{x =  0}{y^{a,b}}{(x+1)^{a,b}}$ 
is dependency-correct: trivially so, since it always propagates all
annotations from the input to each part of the output.  Conversely,
\[G(x^a,y^b) = \left\{\begin{array}{ll}
    y^{a,b} &: x = 0\\
    (x+1)^{a} &: x \neq 0
  \end{array}
\right.\;.
\]
%$G(x^a,y^b) = \ifthenelse{x=0}{y^{a,b}}{(x+1)^{a}}$ 
is dependency-correct and minimally annotated.  To see that $G$ is
dependency-correct, note that if we evaluate $G(x,y)$ on $x \neq 0$
then changing only the value of $y$ (annotated by $b$) can never
change the result.  To see that $G$ is minimally annotated, it suffices
to check that removing any of the annotations breaks
dependency-correctness.
  \end{example}

We say that a query $e$ is \emph{constant} if $\SB{e}\gamma = v$ for
some $v$ and every suitable $\gamma$.  Clearly, a query is constant if
and only if it has a dependency-correct enrichment which annotates
each part of the result with $\emptyset$.

\begin{proposition}
  It is undecidable whether a Boolean NRC query is constant.
\end{proposition}
\begin{proof}
  Recall that query equivalence is undecidable for the relational
  calculus~\cite{ahv}; for NRC, equivalence is undecidable for queries
  $e(x),e'(x)$ over a single variable $x$.  Given two such queries,
  consider the expression $\hat{e} = e(x) \eq e'(x) \orr y$ (definable
  as $\nott(\nott (e(x) \eq e'(x))\andd \nott y)$), where $y$ is a
  fresh variable distinct from $x$.  The result of this expression
  cannot be $\kfalse$ everywhere since the disjunction is $\ktrue$ for
  $y = \ktrue$, so $\hat{e}$ is constant iff $\SB{\hat{e}}\gamma =
  \ktrue$ for every $\gamma$ iff $e \equiv e'$.
\end{proof}
Clearly, an annotation is needed on the result of a Boolean query if
and only if the query is not a constant, so finding minimal
annotations (or minimal slices) is undecidable.  As a result, we
cannot expect to be able to compute minimal dependency-correct
annotations. It is important to note, though, that dependency-tracking
remains hard even if we consider sublanguages for which equivalence is
decidable.  For example, if we just consider Boolean expressions,
finding minimal correct dependency information is also intractable, by an
easy reduction from the validity problem.  These observations motivate
considering approximation techniques, such as those in the next two
sections.

\begin{remark}[Uniqueness of minimum annotations]
  We have shown that annotation minimality is undecidable.  However
  there is another interesting question that we have not answered:
  specifically, given a function $f$, is there a \emph{unique}
  minimally-annotated function $F \enriches f$?  To show this, one
  strategy could be to define a meet (greatest lower bound) operation
  $v \meet w$ on compatible annotated values, lift this to compatible
  annotated functions $(F \meet G)(x) = F(x) \meet G(x)$, show that
  dependency-correctness is preserved by $\meet$, and show that the
  greatest lower bound of the set of all dependency-correct
  enrichments of $f$ exists and is dependency-correct.

  However, actually defining the meet operation on values that
  preserves dependency-correctness appears nontrivial.  For example,
  it does not work to simply define the meet as the pointwise
  intersection of corresponding annotations.  Indeed, this is not even
  well-defined since the ``pointwise intersection'' of
  $\{1^{a},1^{b}\}$ with itself could either be $\{1^{a},1^{b}\}$ or
  $\{1^\emptyset, 1^\emptyset\}$.  We therefore leave the uniqueness
  of minimally annotated functions as a conjecture.
\end{remark}

%\todo[4]{Uniqueness of minimal annotations?}

%--------------------------------------------------------------------%
\section{Dynamic Provenance Tracking}\labelSec{provenance-tracking}
%--------------------------------------------------------------------%

We now consider a \emph{provenance tracking} approach in which we
interpret each expression $e$ as a dependency-correct a-function
$\PP{e}$.  The definition of the provenance-tracking semantics is
shown in \refFig{prov-tracking-sem}.  Auxiliary operations are used to
define $\PP{-}$; these are shown in \refFig{ann-ops}.  In particular,
note that we define an auxiliary operation $(w^\Phi)^{+\Psi} =
w^{\Phi\cup\Psi}$ that adds $\Psi$ to the top-level annotation of an
a-value $w^\Phi$.

% the\todo[8]{Discuss formalism; mention abstract interp and static analysis background}

\begin{figure}[tb]
\[\begin{array}{rclcrcl}
(w^\Phi)^{+\Phi_0} &=& w^{\Phi \cup \Phi_0}
&&
(i_1^{\Phi_1})\hop{+} (i_2^{\Phi_2}) &=& (i_1+i_2)^{\Phi_1 \cup \Phi_2}
\\
\hop{\nott}(b^\Phi) &=& (\nott b)^\Phi
&&
(b_1^{\Phi_1})\hop{\andd} (b_2^{\Phi_2}) &=& (b_1\andd b_2)^{\Phi_1\cup \Phi_2}
\\
\hat{\pi}_i((v_1,v_2)^\Phi) &=& v_i^{+\Phi}
&&
(w_1^{\Phi_1})\hop{\cup} (w_2^{\Phi_2}) &=& (w_1\cup w_2)^{\Phi_1 \cup \Phi_2}
\\
\hat{\kcond}(\ktrue^\Phi,v_1,v_2) &=& v_1^{+\Phi}
&&
\hat{\kcond}(\kfalse^\Phi,v_1,v_2) &=& v_2^{+\Phi}
\end{array}\] 
\begin{eqnarray*}
\hat{\sum}(\{v_1,\ldots,v_n\}^\Phi) &=& (v_1 \hop{+} \cdots \hop{+} v_n)^{+\Phi}\\
\hat{\bigcup}\{v_1,\ldots,v_n\}^\Phi &=& (v_1 \hop{\cup} \cdots \hop{\cup} v_n)^{+\Phi}\\
\{v(x) \mid x \hop{\in} w^\Phi\} &=& \{v(x) \mid x \in w\}^\Phi\\
(w_1^{\Phi_1})\hop{-} (w_2^{\Phi_2}) &=& \{v \in w_1 \mid |v| \not\in |w_2|\}^{\Phi_1\cup \|w_1\|\cup \Phi_2 \cup \|w_2\|}\\
v_1 \hop{\eq} v_2 &=& \left\{\begin{array}{ll}
\ktrue^{\|v_1\| \cup \|v_2\|} & |v_1| = |v_2|\\
\kfalse^{\|v_1\| \cup \|v_2\|} & |v_1| \neq |v_2|
\end{array}\right.
\end{eqnarray*}
\caption{Auxiliary annotation-propagating operations}\labelFig{ann-ops}
\end{figure}
\begin{figure}[tb]
\[\begin{array}{rclcrcl}
  \PP{x}\hg &=& \hg(x)&&
  \PP{\letin{x=e_1}{e_2}}\hg &=& \PP{e_2}(\hg[x \mapsto \PP{e_1}\hg])\\
  \PP{i}\hg &=& i^\emptyset&&
  \PP{e_1 + e_2}\hg &=& (\PP{e_1}\hg) \hop{+} (\PP{e_2}\hg)\\
  \PP{\ksum(e)}\hg &=& \hat{\sum} (\PP{e}\hg)&&
  \PP{b}\hg &=& b^\emptyset\\
  \PP{\nott e}\hg &=& \hat{\nott}(\PP{e}\hg)&&
  \PP{e_1 \andd e_2}\hg &=& (\PP{e_1}\hg) \hop{\andd} (\PP{e_2}\hg)\\
  \PP{(e_1,e_2)}\hg &=& (\PP{e_1}\hg ,\PP{e_2}\hg )^\emptyset&&
  \PP{\pi_i(e)}\hg &=& \hat{\pi_i}(\PP{e}\hg) \quad (i \in \{1,2\})\\
  \PP{\emptyset}\hg &=& \emptyset^\emptyset&&
  \PP{\setof{e}}\hg &=& \{\PP{e}\hg\}^\emptyset\\
  \PP{e_1 \cup e_2}\hg &=& (\PP{e_1}\hg) \hop{\cup} (\PP{e_2}\hg)&&
  \PP{e_1 - e_2}\hg &=& (\PP{e_1}\hg) \hop{-} (\PP{e_2}\hg)\\
  \PP{\bigcup e}\hg &=& \hop{\bigcup}{\PP{e}\hg}&&
  \PP{\{e \mid x \in e_0\}}\hg &=& \{\PP{e}(\hg[x\mapsto v]) \mid v \hop{\in} \PP{e_0}\hg)\}\\
  \PP{e_1 \eq e_2}\hg &=& (\PP{e_1}\hg) \hop{\eq} (\PP{e_2}\hg)
  &&
  \PP{\ifthenelse{e_0}{e_1}{e_2}}\hg &=& \hat{\kcond}(\PP{e_0}\hg,\PP{e_1}\hg,\PP{e_2}\hg)
\end{array}\]
\caption{Provenance-tracking semantics}\labelFig{prov-tracking-sem}
\end{figure}

Many cases involving ordinary programming constructs are
self-explanatory.  Constants always have empty annotations: nothing in
the input can affect them.  Built-in functions such as $+,\andd,\nott$
propagate all annotations on their arguments to the result.  For a
conditional $\ifthenelse{e_0}{e_1}{e_2}$, the result is obtained by
evaluating $e_1$ or $e_2$, and combining the top-level annotation of
the result with that of $e_0$.  A constructed pair has an
empty top-level annotation; in a projection, the top-level annotation
of the pair is merged with that of the returned value.

In the case for let-binding, note that we bind $x$ to the annotated
result of evaluating $e_1$, and then evaluate $e_2$.  It is possible
for the dependencies involved in constructing $x$ to not be propagated
to the result, if $x$ does not happen to be used in evaluating $e_2$.
This safe because query expressions involve neither stateful
side-effects nor nontermination, in contrast to most work on
information flow and slicing in general-purpose languages.  Similarly,
dependencies can be discarded in pair projection expressions
$\pi_i(e)$ and set comprehensions $\{e_2 \mid x \in e_2\}$, and again
this is safe because queries are purely functional and terminating.

%\todo[9]{Explain ``let'' and how dependencies can be discarded}

The cases involving collection types deserve further explanation.  The
empty set is a constant, so has an empty top-level annotation.
Similarly, a singleton set constructor has an empty annotation.  For
union, we take the union of the underlying bags (of annotated values)
and fuse the top-level annotations.  For comprehension, we leave the
top-level annotation alone.  For flattening $\bigcup e$, we take the
lifted union ($\hop{\cup}$) of the elements of $e$ and add the
top-level annotation of $e$.  Similarly, $\hat{\ksum}(e)$ uses
$\hat{+}$ to add together the elements of $e$, fusing their
annotations with that of $e$.  For set difference, to ensure
dependency correctness, we must conservatively include all of the
colors present on either side in the annotation of the top-level
expression.  Similarly, for equality tests, we must include all of the
colors present in either value in the result annotation.

  Note that equivalent expressions $e \equiv e'$ need not satisfy
  $\PP{e} \equiv \PP{e'}$; for example, $x - x \equiv \emptyset$ but
  $\PP{x-x} \not\equiv \emptyset^{\emptyset}$, since if $\hg(x) =
  \{1^d\}^c$ then $\PP{x-x}\hg = \emptyset^{c,d}$.

\begin{figure}[tb]
(a)
\begin{eqnarray*}
\quad \hg &=& [R:=\{(A:1^{a_1},B:1^{b_1}),(A:1^{a_2},B:2^{b_2}),(A:2^{a_3},B:3^{b_3})\},\\
  && S:=\{(C:1^{c_1},D:2^{d_1},E:3^{e_1}),(C:1^{c_2},D:1^{d_2},E:4^{e_2})\}]
\end{eqnarray*}
(b)
\begin{eqnarray*}
\PP{\Pi_A(R)}\hg &=& \{(A:1^{a_{1}}),(A:1^{a_{2}}),(A:2^{a_{3}})\}\\
\PP{\sigma_{A=B}(R)}\hg &=& \{(A:1^{a_1},B:1^{b_1})\}^{a_{123},b_{123}}\\
\PP{R \times S}\hg &= & \{(A:1^{a_1},B:1^{b_1},C:1^{c_1},D:2^{d_1},E:3^{e_1}),\\
&&(A:1^{a_1},B:1^{b_1},C:1^{c_2},D:1^{d_2},E:4^{e_2}),\ldots\}\\
\PP{\Pi_{BE}(\sigma_{A=D}(R\times S))}\hg &= & \{(B:1^{b_1},E:4^{e_2}),(B:2^{b_2},E:4^{e_2}),\\ &&
(B:3^{b_3},E:3^{e_1})\}^{a_{123},d_{12}}\\
\PP{R\cup \rho_{A/C,B/D}(\Pi_{CD}(S))}\hg &=& \{(A:1^{a_1},B:1^{b_1}),(A:1^{a_2},B:2^{b_2}),(A:2^{a_3},B:3^{b_3}),\\ &&
(A:1^{c_1},B:2^{d_1}),(A:1^{c_2},B:1^{d_2})\}\\
\PP{R- \rho_{A/D,B/E}(\Pi_{DE}(S))}\hg &=& \{(A:1^{a_1},B:1^{b_2}),(A:1^{a_2},B:2^{b_2})\}^{a_{123},b_{123},d_{12},e_{12}}\\
\PP{\ksum(\Pi_A(R))}\hg &=& 4^{a_{1},a_2,a_{3}} \\
\PP{\kcount(R)}\hg & = & 3\\
\PP{\kcount(\sigma_{A=B}(R))}\hg &=& 1^{a_{123},b_{123}}\\
\end{eqnarray*}
\caption{(a) Annotated input environment (b) Examples of provenance
  tracking}\labelFig{dynamic-examples}
\end{figure}

\begin{example}
  Consider an annotated input environment $\hg$, shown in
  \refFig{dynamic-examples}(a), of schema $R:\{(A:\kint ,
  B:\kint)\},S:\{(C:\kint, D:\kint, E:\kint)\}$ (we again use
  named-record syntax for readability).  \refFig{dynamic-examples}(b)
  shows the provenance tracking semantics of the example queries from
  \refFig{example-queries}.  We write $a_{123}$ as an abbreviation for
  the set $\{a_1,a_2,a_3\}$, etc.  Note that in the $\kcount$ example
  query, the output depends only on the number of rows in the input
  and not on the field values; we cannot change the number of elements
  of a multiset by changing field values.
\end{example}

\begin{example}[Grouping and aggregation]\labelEx{grouping-ex}
  Consider a query that performs grouping and aggregation, such as
\begin{verbatim}
SELECT A, SUM(B) FROM R GROUP BY A
\end{verbatim}
 First, let
\[X = \{(A:x.A,B:\{y.B \mid y \in R, x.A = y.A\}) \mid x \in R\}\]
When run against the environment $\hg$ in \refFig{dynamic-examples}(a), we obtain result
\[X = \{(A:1^{a_1},B:\{1^{b_1},2^{b_2}\}^{a_{123}}), (A:1^{a_2},B:\{1^{b_1},2^{b_2}\}^{a_{123}}), (A:2^{a_3},B:\{3^{b_3}\}^{a_{123}})\}\]
Note that since we consider collections to be multisets, we get two
copies of $(1,\{1,2\})$, one corresponding to $a_1$ and one
corresponding to $a_2$.  Also, since the subqueries computing the
$B$-values inspect the $A$-values, each of the groups depends on each
of the $A$-values.  We can obtain the final result of aggregation by
evaluating
\begin{eqnarray*}
Y &=& \{(A:x.A, B:\ksum(x.B)) \mid x \in X\}\\
&=& \{(A:1^{a_1},B:3^{a_{123}b_{12}}), (A:1^{a_2},B:3^{a_{123}b_{12}}), (A:2^{a_3},B:3^{a_{123}b_3})\}
\end{eqnarray*}
\end{example}

\begin{remark}
  Our approach to handling negation and equality may result in 
  large annotations in some cases.  For example, consider
  $\{1^a,2^b\}^c - \{1^d,3^e\}^f$. Changing any of the input locations
  $a,b,c,d,e,f$ can cause the output to change.  For example, changing
  $1^a$ to $4^a$ yields result $\{4,2\}$, while changing $2^b$ to
  $3^b$ yields result $\emptyset$.  Thus, we must include all of the colors
  in the input in the annotation of the top-level of the result set,
  since the size of the set can be affected by changes to any of these
  parts.

  Most previous techniques have not attempted to deal with negation.
  One exception is \citeN{DBLP:journals/tods/CuiWW00}'s definition of
  lineage.  In their approach, the lineage of tuple $t \in R - S$
  would be the tuple $t \in R$ and all tuples of $S$.  While
  this is more concise in some cases, it is not dependency-correct by
  our definition.%   Moreover, \citeN{DBLP:journals/tods/CuiWW00}
%   motivated their definition by a semantic criterion that holds if we
%   consider each relational operation (such as selection, projection,
%   join, union, or difference) in isolation, but is not preserved by
%   composition of operators, so does not hold for arbitrary queries and
%   is sensitive to query rewriting.
 
  On the other hand, our approach can also be more concise than
  lineage in the presence of negation, because lineage only deals with
  annotations at the level of records. For example, in $\{1\} -
  \{\pi_1(x) \mid x \in S\}$, our approach will indicate that the
  output does not depend on the second components of elements of $S$,
  whereas the lineage of each tuple in the result of this query
  includes all the records in $S$.  This can make a big difference if
  there are many fields that are never referenced; indeed, some
  scientific databases have tens or hundreds of fields per record,
  only a few of which are needed for most queries.

  Thus, although our approach to negation does exhibit pathological
  behavior in some cases, it also provides more useful provenance for
  other typical queries.  In any case all other approaches either
  ignore negation or also have some pathological behavior.  Developing
  more sophisticated forms of dependence that are better-behaved in
  the presence of negation is an interesting area for future work.
\end{remark}

%--------------------------------------------------------------------%
\subsection{Correctness of dynamic tracking}
%--------------------------------------------------------------------%

In this section, we prove two correctness properties of dynamic
tracking.  First, we show that if $\wf{\Gamma}{e}{\tau}$ then $\PP{e}
: \AAz{\Gamma} \to \AAz{\tau}$ and $\PP{e}\enriches \E{e}$, that is,
the provenance semantics respects the typing and the ordinary
semantics of $e$.  Second, and more importantly, we show that $\PP{e}$
is dependency-correct.  We first establish useful auxiliary properties
of the annotation-merging operation $v^{+\Phi}$ and prove that the
lifted operations such as $\hat{+}$ have appropriate types and enrich
the corresponding ordinary operations.

\begin{lemma}
Let $v$ be an a-value and $\Phi$ an annotation.  Then (1)
$|v^{+\Phi}| = |v|$ and (2) $\|v^{+\Phi}\| = \|v\| \cup \Phi$.
\end{lemma}

\begin{lemma}\labelLem{aux-ops-ok}
  In the following, assume that $v,v_1,v_2$ are in the domains of the
  appropriate functions.
\begin{enumerate}
\item $\hop{+} : \AAz{\kint} \times \AAz{\kint} \to \AAz{\kint}$ is
  color-invariant and $|v_1 \hop{+} v_2| = |v_1| + |v_2|$.
\item $\hat{\sum} : \AAz{\{\kint\}} \to \AAz{\kint}$ is
  color-invariant  and
  $|\hat{\sum}v|= \sum|v|$.
\item $\hat{\nott}: \AAz{\kbool} \to \AAz{\kbool}$ is
  color-invariant and
  $|\hat{\nott}v| = \nott|v|$.
\item $\hat{\andd} : \AAz{\kbool} \times\AAz{\kbool} \to \AAz{\kbool}$ is
  color-invariant 
  and $|v_1 \hop{\andd} v_2| = |v_1| \andd |v_2|$.
\item For any $\tau_1,\tau_2$ and $i \in \{1,2\}$ we have $\hat{\pi}_i
  : \AAz{\tau_1 \times \tau_2} \to \AAz{\tau_i}$ is color-invariant
  and $|\hat{\pi}_i(v)| = \pi_i(|v|)$.
\item For any $\tau$, we have $\hat{\eq} : \AAz{\tau} \times \AAz{\tau} \to
  \AAz{\kbool}$ is
  color-invariant and $|v_1 \hop{\eq} v_2| = (|v_1| \eq |v_2|)$.
\item For any $\tau$, we have $\hat{\kcond} : \AAz{\kbool} \times \AAz{\tau}
  \times \AAz{\tau} \to \AAz{\tau}$ is
  color-invariant and $|\hat{\kcond}(v,v_1,v_2)| =
  \ifthenelse{|v|}{|v_1|}{|v_2|}$.
\item For any $\tau$, we have $\hat{\cup} : \AAz{\{\tau\}}
  \times\AAz{\{\tau\}} \to \AAz{\{\tau\}}$ is
  color-invariant and $|v_1 \hop{\cup} v_2| =
  |v_1| \cup |v_2|$.
\item For any $\tau$, we have $\hat{-} : \AAz{\{\tau\}} \times\AAz{\{\tau\}}
  \to \AAz{\{\tau\}}$ is
  color-invariant and $|v_1 \hop{-} v_2| = |v_1| - |v_2|$.
\item For any $\tau$, we have $\hat{\bigcup} : \AAz{\{\{\tau\}\}} \to
  \AAz{\{\tau\}}$ is color-invariant and $|\hat{\bigcup}v| =
  \bigcup|v|$.
\end{enumerate}
\end{lemma}
\begin{proof*}
  Most cases are immediate.  The cases for sum $(\sum)$ and flattening
  ($\bigcup)$ rely on the cases for binary addition and union.

  The second part of the case of difference (9) is slightly involved.
  We reason as follows.
  \begin{eqnarray*}
    |w_1^{\Phi_1} \hop{-} w_2^{\Phi_2}| &=& |\{v \mid v \in w_1, |v| \not\in |w_2|\}^{\Phi_1\cup \|w_1\| \cup \Phi_2 \cup \|w_2\|}|= \{|v| \mid v \in w_1, |v| \not\in |w_2|\}\\
&=& \{v \mid v \in |w_1|, v \not\in |w_2|\}= |w_1| - |w_2|
  \end{eqnarray*}
\end{proof*}

\begin{lemma}\labelLem{prov-track-ok}
  If $\wf{\Gamma}{e}{\tau}$ then $\PP{e} : \AAz{\Gamma} \to
  \AAz{\tau}$ is color-invariant and $\PP{e} \enriches \E{e}$.
\end{lemma}
\begin{proof}
  Proof is by induction on expressions $e$ (which determine the
  structure of the typing judgment).  Most cases are straightforward,
  given \refLem{aux-ops-ok}; we show the case of comprehensions.
\begin{itemize}
  \item Case $e =\{e_2 \mid x \in e_1\}$:
\[
\infer{\wf{\Gamma}{\{e_2 \mid x \in e_1\}}{\setof{\tau_2}}}
{\wf{\Gamma}{e_1}{\setof{\tau_1}} & \wf{\Gamma,x{:}\tau_1}{e_2}{\tau_2}}
\]

First, by induction we have $\PP{e_1}: \AAz{\Gamma} \to
\AAz{\{\tau_1\}}$.  Hence $w^\Phi := \PP{e_1}\hg$
is a set of a-values in $\AAz{\tau_1}$.  So for each $v \hop{\in}
\PP{e_1}\hg$, we have $\hg' := \hg[x:=v] \in \AAz{\Gamma,x{:}\tau_1}$,
hence $\PP{e_2}\hg' \in \AAz{\tau_2}$, and so
\[\PP{\{e_2 \mid x \in e_1\}} \hg 
= \{\PP{e_2}\hg[x:=v] \mid v \in w\}^{\Phi} \in \AAz{\{\tau_2\}}\]
Furthermore, if $\alpha : \kcolor \to \{\kcolor\}$, then we have
\begin{eqnarray*}
\alpha(\PP{\{e_2 \mid x \in e_1\}} \hg) 
&=& \alpha(\{\PP{e_2}(\hg[x:=v]) \mid v \in w\}^{\Phi}) \\
&=& \{\alpha(\PP{e_2}(\hg[x:=v])) \mid v \in w\}^{\alpha[\Phi]} \\
&=& \{\PP{e_2}(\alpha(\hg)[x:=\alpha(v)]) \mid v \in w\}^{\alpha[\Phi]} \\
&=& \{\PP{e_2}(\alpha(\hg)[x:=v]) \mid v \in \alpha(w)\}^{\alpha[\Phi]}\\
&=& \{\PP{e_2}(\alpha(\hg)[x:=v]) \mid v \hop{\in} \alpha(w)^{\alpha[\Phi]}\}\\
&=& \{\PP{e_2}(\alpha(\hg)[x:=v]) \mid v \hop{\in} \alpha(\PP{e_1}\hg)\}\\
&=& \{\PP{e_2}(\alpha(\hg)[x:=v]) \mid v \hop{\in} \PP{e_1}\alpha(\hg)\}\\
&=& \PP{\{e_2 \mid x \in e_1\}} \alpha(\hg)
\end{eqnarray*} 
where we appeal to the induction hypothesis to show that $\PP{e_1}$
and $\PP{e_2}$ are color-invariant.  Hence $\PP{\{e_2 \mid x \in
  e_1\}}$ is color-invariant.

Second, to show that $\PP{\{e_2 \mid x \in e_1\}} \enriches \E{\{e_2
  \mid x \in e_1\}}$, we have:
\begin{eqnarray*}
 \E{\{e_2\mid x \in e_1\}}|\hg| 
&=& \{\E{e_2}|(\hg|[x:=v]) \mid v \in \E{e_1}|\hg|\}
= \{\E{e_2}|(\hg|[x:=v]) \mid v \in |\PP{e_1}\hg|\}\\
&=& \{\E{e_2}|(\hg|[x:=v]) \mid v \in |w^\Phi|\}
= \{\E{e_2}|(\hg|[x:=|v|]) \mid |v| \in |w^\Phi|\}\\
&=& \{\E{e_2}|(\hg|[x:=|v|]) \mid v \in w\}
= \{\E{e_2}|\hg[x:=v]| \mid v \in w\}\\
&=& \{|\PP{e_2}(\hg[x:=v])| \mid v \in w\}
= |\{\PP{e_2}(\hg[x:=v]) \mid v \in w\}^\Phi |\\
&=& |\{\PP{e_2}(\hg[x:=v]) \mid v \hop{\in} w^\Phi\} |
= |\{\PP{e_2}(\hg[x:=v]) \mid v \hop{\in} \PP{e_1}\hg\} |\\
&=& |\PP{\{e_2 \mid x \in e_1\}} \hg |
\end{eqnarray*}
  \end{itemize}
\end{proof}

We now turn to dependency-correctness.  
Since $\PP{e}$ is defined in terms of the special
annotation-propagating operations introduced in \refFig{ann-ops}, we need
to show that these operations are dependency-correct.  We first need
to establish properties of $\equiv_a$:

\begin{lemma}\labelLem{a-equiv}
\begin{enumerate}
\item If $v \equiv_a v'$ then $a \in \|v\| \iff a \in \|v'\|$.
\item If $a \not\in \|v\|$ and $v \equiv_a v'$ then $v = v'$.
\item If $v_1 \equiv_a v_2$ then $v_1^{+\Phi} \equiv_a v_2^{+\Phi}$.
\end{enumerate}
\end{lemma}
\begin{proof}
  The first part is easy to establish by induction on derivations of
  $\equiv_a$, by noting that $a \in \|v\| \iff a \in \|v'\|$ is
  equivalent to $\|v \|\cap \{a\} = \|v'\| \cap \{a\}$ and reasoning
  equationally.

For the second part, note that the rule
\[
\infer{w_1^{\Phi_1} \equiv_a w_2^{\Phi_2}}{w_1 \equiv_a w_2 & a \in \Phi_1 \cap \Phi_2}
\]
can never apply since $a \notin \|w_1^{\Phi_1}\| = \|w_1\| \cup
\Phi_1$ implies $a \not\in \Phi_1 \cap \Phi_2$. The remaining
rules coincide with the rules for annotated value equality.

For the third part observe that both of the rules defining $\equiv_a$ for
annotated values are preserved by adding equal sets of annotations to
both sides.
\end{proof}

We now state a key lemma which shows that all of the lifted operations
are dependency-correct.  Many of the arguments are similar.  In each
case, if we know that the inputs to an operation are $\equiv_a$, we
reason by cases on the structure of the derivation of $\equiv_a$.  If
any of the assumptions $v\equiv_a v'$ hold because $a \in \Phi \cap \Phi'$ for
some pair of inputs $v,v'$, then both outputs will also be annotated
with $a$. Otherwise, the inputs must have the same top-level
structure, so in each case we have enough information to evaluate the
unlifted function and show that the results are still $\equiv_a$.

The proofs for equality and difference operations are slightly
different.  Both operations are potentially global, that is,
changes deep in the input values can affect the top-level structure of
the result (trivially for $\eq$, since there is no deep structure in
the boolean result).  This is, essentially, why we need to include all
of the annotations of the inputs in the result of an equality or
difference operation.  We should point out that this inaccuracy is an
area where we believe improvement may be possible, through refining
the definition of $\equiv_a$; but this is left for future work.

\begin{lemma}\labelLem{lifted-dependency-correct}
If $v \equiv_a v'$, $v_1 \equiv_a v_1',v_2 \equiv v_2',\ldots$ then:
\begin{enumerate}
  \item $v_1 \hop{+} v_2 \equiv_a v_1' \hop{+} v_2'$
  \item $\hat{\sum}v \equiv_a \hat{\sum}v'$
  \item $\hat{\nott}v \equiv_a \hat{\nott}v'$
  \item $v_1 \hop{\andd} v_2 \equiv_a v_1' \hop{\andd} v_2'$
  \item $\hat{\pi}_i(v) \equiv_a \hat{\pi}_i(v')$
  \item $v_1 \hop{\eq} v_2 \equiv_a v_1' \hop{\eq} v_2'$
  \item $\hat{\kcond}(v,v_1,v_2) \equiv_a \hat{\kcond}(v',v_1',v_2')$
  \item $v_1 \hop{\cup} v_2 \equiv_a v_1' \hop{\cup} v_2'$
  \item $v_1 \hop{-} v_2 \equiv_a v_1' \hop{-} v_2'$
  \item $\hat{\bigcup}v \equiv_a \hat{\bigcup}v'$
\end{enumerate}
\end{lemma}
\begin{proof}
  For part (1), suppose $v_i = n_i^{\Phi_i}$ and $v_i' = m_i^{\Psi_i}$
  for $i \in \{1,2\}$.  There are four cases, depending on the
  derivations of $n_i\equiv_a m_i$ for $i \in \{1,2\}$.  If both
  derivations follow because $n_i^{\Phi_i} = m_i^{\Psi_i}$ then
  $\PP{e}\hg = (n_1+n_2)^{\Phi_1\cup\Phi_2} =
  (m_1+m_2)^{\Psi_1\cup\Psi_2} = \PP{e}\hg'$ so again $\PP{e}\hg
  \equiv_a \PP{e}\hg'$.  Otherwise one or both of the derivations
  follows because $a \in \Phi_i \cap \Psi_i$ for $i =1$ or $i=2$.
  Then $a \in (\Phi_1\cup\Phi_2) \cap (\Psi_1\cup\Psi_2)$ so again
  $\PP{e}\hg \equiv_a \PP{e}\hg'$.

  For part (2), there are two cases.  If the summed sets are
  $\equiv_a$ because their top-level annotations mention $a$, then the
  results of the sums will also mention $a$, so we are done.
  Otherwise, we must have that the summed sets are of equal size and
  their elements are pairwise matched by $\equiv_a$; hence, we can
  apply part (1) repeatedly (and then
  \refLem{lifted-dependency-correct}) to show that the results are
  $\equiv_a$.

Parts (3,4) are similar to part (1).

For part (5), suppose $v = (v_1,v_2)^\Phi$ and $v' =
(v_1',v_2')^{\Phi'}$.  Note that \[\hat{\pi_i}(v) = \hat{\pi_i}
(v_1,v_2)^\Phi = v_i^{+\Phi}\] and similarly $\hat{\pi_i}(v') =
(v_i')^{+\Phi'}$.  There are two cases depending on the last step in
the derivation of $v \equiv_a v'$.  If $a \in \Phi \cap \Phi'$ then we
are done since $a$ will be in the top-level annotations of both
$v_i^{+\Phi}$ and $(v_i')^{+\Phi'}$.  Otherwise we must have $v_i
\equiv_a v_i'$ for $i \in \{1,2\}$, so again $v_i^{+\Phi} \equiv_a
(v_i')^{+\Phi'}$.

For part (6), there are two cases.  If $a \in (\|v_1\| \cup \|v_2\|)
\cap (\|v_1'\| \cup \|v_2'\|)$ then we are done.  Otherwise by
\refLem{a-equiv}, $a$ cannot appear anywhere in $v_1,v_2,v_1',v_2'$,
so we must have $v_1 = v_1',v_2 = v_2'$.  Hence $(v_1 \hop{\eq} v_2) =
(v_1' \hop{\eq} v_2')$ which implies the two sides are $\equiv_a$ as
well.

For part (7), suppose $v = b^\Phi, v' = (b')^{\Phi'}$.  If $a \in \Phi
\cap \Phi'$ then we are done since both conditionals will have $a$ in
their top-level annotation.  Otherwise we must have $b = b'$ so
$\hat{\kcond}(v,v_1,v_2) = v_i$ and $\hat{\kcond}(v',v_1',v_2') =
v_i'$, so by induction (and \refLem{a-equiv}) we are done.

For part (8), suppose $v_i = w_i^{\Phi_i}$ and similarly for $v_i'$.
Again if $a \in (\Phi_1 \cup \Phi_2) \cap (\Phi_1' \cup \Phi_2')$ then
we are done.  Otherwise we must have that $w_1 =
\{v_{11},\ldots,v_{1n}\}$, $w_1' = \{v_{11}',\ldots,v_{1n}'\}$ where
$v_{1i} \equiv_a v_{1i}'$ for each $i \in \{1,\ldots,n\}$, and
similarly for $w_2,w_2'$.  Hence the elements of the union of the two
multisets can be matched up using the $\equiv_a$ relation, so we can
conclude that $w_1 \cup w_2\equiv_a w_1' \cup w_2'$ as well.  We must
also have $\Phi_i = \Phi_i'$ for each $i \in \{1,2\}$, so we can
conclude that
\[v_1 \cup v_2 =
(w_1 \cup w_2)^{\Phi_1 \cup \Phi_2}\equiv_a (w_1' \cup w_2')^{\Phi_1'
  \cup \Phi_2'} = v_1' \cup v_2'\]

For part (9), the reasoning is similar to part (6).

For part (10), the reasoning is similar to that for part (2),
appealing to part (8) once we have expanded to binary unions.
\end{proof}

We conclude the section with the proof of dependency-correctness.  It
is much simplified by the previous lemma, since many cases now consist
only of applying the induction hypothesis and then using
dependency-correctness of a lifted operation.

\begin{theorem}\labelThm{dep-correctness}
  If $\wf{\Gamma}{e}{\tau}$ then $\PP{e}$ is dependency-correct.
\end{theorem}
\begin{proof}
  Suppose $\hg \equiv_a \hg'$.  Again proof is by induction on the
  structure of expressions/typing derivations.  Many cases are
  immediate using the induction hypothesis and the corresponding parts
  of \refLem{lifted-dependency-correct}.  We show the remaining cases:
  \begin{itemize}
  \item Case $e = x$: 
\[
\infer{\wf{\Gamma}{x}{\tau}}{x{:}\tau \in \Gamma}
\]
By assumption $\PP{x}\hg = \hg(x) \equiv_a \hg'(x) = \PP{x}\hg'$.

  \item Case $e = \letin{x=e_1}{e_2}$: 
\[
\infer{\wf{\Gamma}{\letin{x=e_1}{e_2}}{\tau_2}}
{\wf{\Gamma}{e_1}{\tau_1} & \wf{\Gamma,x{:}\tau_1}{e_2}{\tau_2}}\] 
By induction $\PP{e_1}$ and $\PP{e_2}$ are dependency-correct.  Hence
$\PP{e_1}\hg \equiv_a \PP{e_1}\hg'$, so $\hg[x:=\PP{e_1}\hg] \equiv_a
\hg'[x:=\PP{e_1}\hg']$.  It then follows by induction that $\PP{e}\hg
= \PP{e_2} (\hg[x:=\PP{e_1}\hg]) \equiv_a
\PP{e_2}(\hg'[x:=\PP{e_1}\hg']) = \PP{e}\hg'$.

  \item Case $e=(e_1,e_2)$:
\[
\infer{\wf{\Gamma}{(e_1,e_2)}{\tau_1 \times \tau_2}}
{
\wf{\Gamma}{e_1}{\tau_1}
&
\wf{\Gamma}{e_2}{\tau_2}
}
\]
By induction, $\PP{e_1}$ and $\PP{e_2}$ are dependency-correct, so
$v_i = \PP{e_i}\hg \equiv_a \PP{e_i}\hg' = v_i'$ for $i \in \{1,2\}$.
Hence we can immediately derive $(v_1,v_2)^\emptyset \equiv_a
(v_1',v_2')^\emptyset$.

  \item Case $e =\{e'\}$:  
\[
\infer{\wf{\Gamma}{\setof{e'}}{\setof{\tau}}}{\wf{\Gamma}{e'}{\tau}}
\]
By induction, $\PP{e'}$ is dependency-correct, so
$v = \PP{e'}\hg \equiv_a \PP{e'}\hg' = v'$.  Hence we can
immediately derive $\{v\}^\emptyset \equiv_a
\{v'\}^\emptyset$.

  \item Case $e =\{e_2 \mid x \in e_1\}$:
\[
\infer{\wf{\Gamma}{\{e_2 \mid x \in e_1\}}{\setof{\tau_2}}}
{\wf{\Gamma}{e_1}{\setof{\tau_1}} & \wf{\Gamma,x{:}\tau_1}{e_2}{\tau_2}}
\]
By induction, $\PP{e_1}$ and $\PP{e_2}$ are dependency-correct.  Hence
$w_1^{\Phi_1} = \PP{e_1}\hg \equiv_a \PP{e_1}\hg' = (w_1')^{\Phi_1'}$.
There are two cases.  If $a \in \Phi_1 \cap \Phi_1'$ then we are done
since $\PP{e}\hg$ and $\PP{e}\hg'$ will both contain top-level
annotations $a$.  Otherwise, we must have
\[
\infer{w_1^{\Phi_1} \equiv_a (w_1')^{\Phi_1'}}{\Phi_1 = \Phi_1' & \infer{ w_1 \equiv_a w_1'}{v_{11} \equiv_a v_{11}' & \cdots & v_{1n} \equiv_a v_{1n}'}}
\]
where $w_1 = \{v_{11},\ldots,v_{1n}\}$ and similarly for $w_1'$.
Thus, for each $i \in \{1,\ldots,n\}$, we have $\hg[x:=v_{1i}]
\equiv_a \hg'[x:=v_{1i}']$.  It follows that for some $v_{2i}$ and
$v_{2i}'$, we have $v_{2i} = \PP{e_2}(\hg[x:=v_{1i}]) \equiv_a
\PP{e_2}(\hg[x:=v_{1i}'])= v_{2i}' $ for each $i \in \{1,\ldots,n\}$.
Thus, we can derive
\[
\infer{w_2^{\Phi_1} \equiv_a (w_2')^{\Phi_1'}}{\Phi_1 = \Phi_1' & \infer{ w_2 \equiv_a w_2'}{v_{21} \equiv_a v_{21}' & \cdots & v_{2n} \equiv_a v_{2n}'}}
\]
where 
\[w_2^{\Phi_1} = \{\PP{e_2}(\hg[x:=v]) \mid v \in w_1\}^{\Phi_1} % =
% \{\PP{e_2}(\hg[x:=v]) \mid v \hop{\in} \PP{e_1}\hg\}
= \PP{\{e_2\mid x
  \in e_1\}}\hg\]
and similarly 
\[(w_2')^{\Phi_1'} = \{\PP{e_2}(\hg'[x:=v]) \mid v \in w_1\}^{\Phi_1} =
%\{\PP{e_2}(\hg'[x:=v]) \mid v \hop{\in} \PP{e_1}\hg'\} = 
\PP{\{e_2\mid x
  \in e_1\}}\hg'\]
So
we can conclude that $\PP{\{e_2\mid x \in e_1\}}\hg \equiv_a
\PP{\{e_2\mid x \in e_1\}}\hg'$.
  \end{itemize}
This exhausts all cases and completes the proof.
\end{proof}

%--------------------------------------------------------------------%
\section{Static Provenance Analysis}\labelSec{provenance-analysis}
%--------------------------------------------------------------------%

Dynamic provenance may be expensive to compute and nontrivial to
implement in a standard relational database system.  Moreover, dynamic
analysis cannot tell us anything about a query without looking at
(annotated) input data.  In a typical large database, most of the data
is in secondary storage, so it is worthwhile to be able to avoid data
access whenever possible.  Moreover, even if we want to perform
dynamic provenance tracking, a static approximation of dependency
information may be useful for optimization.  In this section we
consider a \emph{static provenance analysis} which statically
approximates the dynamic provenance, but can be calculated quickly
without accessing the input.

We formulate the analysis as a type-based
analysis\cite{palsberg01paste}; annotated types (a-types) $\htau$ and
raw types (r-types) $\omega$ are defined as follows:
\begin{eqnarray*}
\htau &::=& \omega^\Phi \qquad \omega ::= \intTy \mid \boolTy \mid \htau \times \htau' \mid \{\htau\}
\end{eqnarray*}
We write $\hG$ for a typing context mapping variables to a-types.
We lift the auxiliary a-value operations of erasure ($|\htau|$) and
annotation extraction ($\|\htau\|$) to a-types as follows:
\[\begin{array}{rcl}
|\intTy| &=& \intTy\\
|\boolTy| &=& \boolTy\\
|\htau_1 \times \htau_2| &=& |\htau_1| \times |\htau_2|\\
|\{\htau\}| &=& \{|\htau|\}\\
|\omega^\Phi| &=& |\omega|
\end{array}
\quad
\begin{array}{rcl}
\|\intTy\| &=& \emptyset\\
\|\boolTy\| &=& \emptyset\\
\|\htau_1 \times \htau_2\| &=& \|\htau_1\| \cup \|\htau_2\|\\
\|\{\htau\}\| &=& \|\htau\|\\
\|\omega^\Phi\| &=& \|\omega\| \cup \Phi
\end{array}
\]
Moreover, we define compatibility for a-types analogously to
compatibility for values, that is, $\htau_1$ and $\htau_2$ are
compatible ($\htau_1 \cong \htau_2$) provided $|\htau_1| = |\htau_2|$.
Also, we say that an a-type \emph{enriches} an ordinary type $\tau$
(written $\htau \enriches \tau$) provided $|\htau| = \tau$. These concepts
are lifted to a-contexts $\hG$ mapping variables to types in the
obvious (pointwise) way.

We also define a merge
operation $\sqcup$ on compatible types as follows:
\[\begin{array}{rcl}
\intTy \sqcup \intTy &=& \intTy
\\
\boolTy \sqcup \boolTy &=& \boolTy
\\
(\htau_1 \times \htau_2) \sqcup (\htau_1' \times \htau_2') &=& (\htau_1 \sqcup \htau_1') \times (\htau_2 \sqcup \htau_2')
\\
\{\htau\} \sqcup \{\htau'\} &=& \{\htau\sqcup \htau'\}
\\
\omega_1^{\Phi_1} \sqcup \omega_2^{\Phi_2} &=& (\omega_1\sqcup \omega_2)^{\Phi_1 \cup \Phi_2} 
\end{array}\]
Finally, we write $\htau \sqsubseteq \htau'$ if $\htau' = \htau \sqcup
\htau'$; this is a partial order on types and can be viewed as a
subtyping relation.

We interpret a-types $\htau$ as sets of a-values $\AA{\htau}$.  We
interpret the annotations in a-types as upper bounds on the
annotations in the corresponding a-values:
\[\begin{array}{rcl}
\AA{\intTy} &=& \{i \mid i \in \Int\}
\\
  \AA{\boolTy} &=& \{b \mid b \in \Bool\}
\\
\AA{\htau_1\times\htau_2} &=& \AA{\htau_1} \times  \AA{\htau_2}
\\
  \AA{\setof{\htau}} &=& \mathcal{M}_\kfin(\AA{\htau})
\\
\AA{\omega^\Phi} &=& \{w^{\Psi} \mid \Psi \subseteq \Phi, w \in \AA{\omega}\}
\end{array}\]
The syntactic operations $|{-}|$, $\|{-}\|$, $\sqsubseteq$ and
$\sqcup$ on types correspond to appropriate semantic operations on
sets of a-values.  We note some useful properties of these operations:
\begin{lemma}
\begin{enumerate}
\item If $v \in \AA{\htau}$ then $v \in \AAz{|\htau|}$ and $|v| \in
  \T{|\htau|}$ and $\|v\| \subseteq \|\htau\|$.
\item If $\htau_1 \cong \htau_2$ then $\htau_1 \sqcup \htau_2$ is
  defined and $\AA{\htau_1 \sqcup \htau_2} \supseteq \AA{\htau_1} \cup
  \AA{\htau_2}$ and $\|\htau_1 \sqcup \htau_2\| = \|\htau_1\| \cup
  \|\htau_2\|$.
\item If $\htau_1 \sqsubseteq \htau_2$ then $\AA{\htau_1} \subseteq
  \AA{\htau_2}$ and $\|\htau_1\| \subseteq \|\htau_2\|$.
\end{enumerate}
\end{lemma}

\begin{figure}[tb]
\[\begin{array}{c}
\infer{\wf{\hG}{x}{\htau}}{x{:}\htau \in \hG} 
\quad
\infer{\wf{\hG}{\letin{x=e_1}{e_2}}{\htau_2}}
{
\wf{\hG}{e_1}{\htau_1}
& 
\wf{\hG,x:\htau_1}{e_2}{\htau_2}
}
\smallskip\\
\infer{\wfeff{\hG}{i}{\intTy }{ \emptyset}}{}
\quad
\infer{\wfeff{\hG}{e_1+e_2}{\intTy}{\Phi_1 \cup \Phi_2}}
{\wfeff{\hG}{e_1}{\intTy}{\Phi_1}
& 
\wfeff{\hG}{e_2}{\intTy}{\Phi_2}}
\quad
\infer{\wfeff{\hG}{\ksum(e)}{\intTy}{\Phi_0 \cup \Phi}}
{\wfeff{\hG}{e}{\setof{\intTy^{\Phi_0}}}{\Phi}}
\smallskip\\
\infer{\wfeff{\hG}{b}{\boolTy }{ \emptyset}}{}
\quad
\infer{\wfeff{\hG}{\neg e}{\boolTy}{\Phi}}{\wfeff{\hG}{e}{\boolTy}{\Phi}}
\quad
\infer{\wfeff{\hG}{e_1 \andd e_2}{\boolTy}{\Phi_1 \cup\Phi_2}}{\wfeff{\hG}{e_1}{\boolTy}{\Phi_1}&\wfeff{\hG}{ e_2}{\boolTy}{\Phi_2}}
\smallskip\\
\infer{\wfeff{\hG}{(e_1,e_2)}{(\htau_1 \times \htau_2)}{\emptyset}}
{
\wf{\hG}{e_1}{\htau_1}
&
\wf{\hG}{e_2}{\htau_2}
}
\quad
\infer[(i \in \{1,2\})]{\wfeff{\hG}{\pi_i(e)}{\omega_i}{\Phi_i \cup \Phi}}{
\wfeff{\hG}{e}{\omega_1^{\Phi_1} \times \omega_2^{\Phi_2}}{\Phi}
}
\smallskip\\
\infer{\wfeff{\hG}{e_1 \eq e_2}{\boolTy}{\|\htau_1\| \cup \|\htau_2\|}}{
\wf{\hG}{e_1}{\htau_1}
&
\wf{\hG}{ e_2}{\htau_2} 
& 
\htau_1 \cong \htau_2}
\quad
\infer{\wf{\hG}{\ifthenelse{e_0}{e_1}{e_2}}{(\htau_1 \sqcup \htau_2)^{+\Phi_0}}}
{\wfeff{\hG}{e_0}{\kbool}{\Phi_0}
&
\wf{\hG}{e_1}{\htau_1}
& 
\wf{\hG}{e_2}{\htau_2}
& \htau_1 \cong \htau_2}
\smallskip\\
\infer{\wfeff{\hG}{\emptyset}{\{\htau\} }{ \emptyset}}{}
\quad
\infer{\wfeff{\hG}{\setof{e}}{\setof{\htau}}{\emptyset}}{\wf{\hG}{e}{\htau}}
\quad
\infer{\wfeff{\hG}{e_1 \cup e_2}{\setof{\htau_1 \sqcup \htau_2}}{\Phi_1 \cup \Phi_2}}
{\wfeff{\hG}{e_1 }{\setof{\htau_1}}{\Phi_1}& \wfeff{\hG}{e_2 }{\setof{\htau_2}}{\Phi_2} & \htau_1 \cong \htau_2}
\smallskip\\ 
\infer{\wfeff{\hG}{\{e_2 \mid x \in e_1\}}{\setof{\omega^{\Phi_2}}}{\Phi_1}}
{
\wfeff{\hG}{e_1}{\setof{\htau_1} }{ \Phi_1} 
& 
\wfeff{\hG,x{:}\htau_1}{e_2}{\omega}{\Phi_2}
}
\quad
\infer{\wfeff{\hG}{\bigcup{e}}{\setof{\htau}}{\Phi_1 \cup \Phi_2}}
{ \wfeff{\hG}{e}{\setof{\setof{\htau}^{\Phi_2}}}{\Phi_1}}
\smallskip\\
\infer{\wfeff{\hG}{e_1 - e_2}{\setof{\htau_1}}{\|\{\htau_1\}^{\Phi_1}\| \cup \|\{\htau_2\}^{\Phi_2}\|}}
{\wfeff{\hG}{e_1 }{\setof{\htau_1}}{\Phi_1}& \wfeff{\hG}{e_2 }{\setof{\htau_2}}{\Phi_2}& \htau_1 \cong \htau_2}
\end{array}\]
\caption{Type-based static provenance analysis}\labelFig{approximate-prov}
\end{figure}

\refFig{approximate-prov} shows the annotated typing judgment
$\wf{\hG}{e}{\htau}$ (sometimes written $\wfeff{\hG}{e}{\omega}{\Phi}$
for readability, provided $\htau = \omega^\Phi$), which extends the
plain typing judgment shown in \refFig{expression-typing}.
\begin{proposition}
  The judgment $\wf{\Gamma}{e}{\tau}$ is derivable if and only if for
  any $\hG \enriches \Gamma$, there exists a $\htau \enriches \tau$ such that $\wf{\hG}{e}{\htau}$.  Moreover, given $\Gamma
  \vdash e : \tau$ and $\hG \enriches\Gamma$, we can compute
  $\htau$ in polynomial time  (by a simple syntax-directed
algorithm).
\end{proposition}

\begin{example}
  Consider an annotated type context $\hG$, shown in
  \refFig{static-examples}(a), where we have annotated field values
  $A,B,C,D,E$ with colors $a,b,c,d,e$ respectively.
  \refFig{static-examples}(b) shows the results of static analysis for
  the queries in \refFig{dynamic-examples}.  In some cases, the type
  information simply reflects the field names which are present in the
  output.  However, the colors are not affected by renamings, as in
  $\rho_{A/C,B/D}$.  Furthermore, note that (if we replace the colors
  $a,b,c,d,e$ with color sets $\{a_1,a_2,a_3\}$, etc.) in each case
  the type-level colors safely over-approximate the value-level colors
  calculated in \refFig{dynamic-examples}.
\end{example}

\begin{figure}[tb]
(a)
\[\hG = [R:\{(A:\kint^{a} , B:\kint^{b})\},S:\{(C:\kint^{c},D:\kint^{d},E:\kint^{e})\}]\]
\\
(b)
\[\begin{array}{lcl}
\wf{\hG}{\Pi_A(R)&}{&\{(A:\kint^{a})\}} \smallskip\\
\wf{\hG}{\sigma_{A=B}(R)&}{&\{(A:\kint^{a},B:\kint^{b})\}^{a,b}}\smallskip\\
\wf{\hG}{R \times S&}{&\{(A:\kint^{a},B:\kint^{b},C:\kint^{c},D:\kint^{d},E:\kint^{e})\}}\smallskip\\
\wf{\hG}{\Pi_{BE}(\sigma_{A=D}(R\times S))&}{&\{(B:\kint^{b},E:\kint^{e})\}^{a,d}}\smallskip\\
\wf{\hG}{R\cup \rho_{A/C,B/D}(\Pi_{CD}(S))&}{&\{(A:\kint^{a,c},B:\kint^{b,d})\}}\smallskip\\
\wf{\hG}{R- \rho_{A/D,B/E}(\Pi_{DE}(S))&}{&\{(A:\kint^{a},B:\kint^{b})\}^{a,b,d,e}}\smallskip\\
\wf{\hG}{\ksum(\Pi_A(R))&}{&\kint^{a}}\smallskip\\
\wf{\hG}{\kcount(R)&}{&\kint}\smallskip\\
\wf{\hG}{\kcount(\sigma_{A=B}(R))&}{&\kint^{a,b}}
\end{array}\]
\caption{(a) Annotated input context (b) Examples of provenance
  analysis}\labelFig{static-examples}
\end{figure}

\begin{example}
  To further illustrate the analysis, we consider an
  extended example for a query that performs grouping and aggregation
  (equivalent to the one in \refEx{grouping-ex}):
  \[Q(R) = \{(\pi_1(x),\ksum(G(x))) \mid x \in R\}\]
  where we employ the following abbreviations:
  \begin{eqnarray*}
    G(x) &:=& \bigcup\{\ifthenelse{\pi_1(y)\eq\pi_1(x)}{\{\pi_2(y)\}}{\emptyset} \mid y \in R\}\\
    \htau_R &:=& \kint^{a} \times \kint^{b}\\
    \hG &:=& R{:}\{\htau_R\}\\
    \hG_1 &:=& \hG,x{:}\htau_R\\
    \hG_2 &:=& \hG_1,y{:}\htau_R
\end{eqnarray*}
  We will derive $\wf{\hG}{Q(R)}{\{\kint^a \times \kint^{a,b}\}}$.
  The derivation illustrates how color $a$ is propagated to both parts
  of the result type, while color $b$ is only propagated to the second
  column.
  
  First, we can reduce the analysis of $Q$ to analyzing $G(x)$ as
  follows:
  \[
  \begin{array}{c}
    \infer{\wf{\hG}{\{(\pi_1(x),\ksum(G(x))) \mid x \in R\}}{\{\kint^a\times \kint^{a,b}\}}}
    {\wf{\hG}{R}{\{\htau_R\}} & 
      \infer{\wf{\hG_1}{(\pi_1(x),\ksum (G(x)))}{\kint^a\times \kint^{a,b}}}
      {\infer{\wf{\hG_1}{\pi_1(x)}{\kint^a}}
        {\wf{\hG_1}{x}{\htau_R}} & 
        \infer{\wf{\hG_1}{\ksum (G(x))}{\kint^{a,b}}}{
          \wf{\hG_1}{G(x)}{\{\kint^b\}^a}
        }
      }
    }\end{array}
  \]
  We next reduce the analysis of $G(x)$ to an analysis of the conditional
  inside $G(x)$:
  \[
  \begin{array}{c}
    \infer{\wf{\hG_1}{\bigcup\{\ifthenelse{\pi_1(y)\eq\pi_1(x)}{\{\pi_2(y)\}}{\emptyset} \mid y \in R\}}{\{\kint^b\}^a}}{
      \infer{\wf{\hG_1}{\{\ifthenelse{\pi_1(y)\eq\pi_1(x)}{\{\pi_2(y)\}}{\emptyset} \mid y \in R\}}{\{\{\kint^b\}^a\}}}{
        \wf{\hG_1}{R}{\{\htau_R\}} 
        & 
        \wf{\hG_2}{\ifthenelse{\pi_1(y)\eq\pi_1(x)}{\{\pi_2(y)\}}{\emptyset}}{\{\kint^b\}^a}
      }
    }
  \end{array}
  \]
  Finally, we can analyze the conditional as follows:
  \[
  \begin{array}{c}
    \infer{\wf{\hG_2}{\ifthenelse{\pi_1(y)\eq\pi_1(x)}{\{\pi_2(y)\}}{\emptyset}}{\{\kint^{b}\}^a}}{
      \infer{\wf{\hG_2}{\pi_1(y)\eq\pi_1(x)}{\kbool^a}}{
        \infer{\wf{\hG_2}{\pi_1(y)}{\kint^a}}{
          \infer{\wf{\hG_2}{y}{\htau_R}}{
          }
        }
        &
        \infer{\wf{\hG_2}{\pi_1(x)}{\kint^a}}{
          \infer{\wf{\hG_2}{x}{\htau_R}}{
          }
        }
      }
      & 
      \infer{\wf{\hG_2}{\{\pi_2(y)\}}{\{\kint^b\}}}{
        \infer{\wf{\hG_2}{\pi_2(y)}{\kint^b}}{
          \infer{\wf{\hG_2}{y}{\htau_R}}{
          }
        }
      }
      & 
      \infer{\wf{\hG_2}{\emptyset}{\{\kint\}}}{
      }
    }
\end{array}
\]
\end{example}

%\todo[10]{Discuss introduction example again here?}

%--------------------------------------------------------------------%
\subsection{Correctness of static analysis}
%--------------------------------------------------------------------%
The correctness of the analysis is proved with respect to the
provenance-tracking semantics given in \refSec{provenance-tracking},
which we have already shown dependency-correct.  Correctness is
formulated as a type-soundness theorem, using the refined
interpretation $\AA{-}$ of a-types.  Specifically, we show that if
$\wf{\hG}{e}{\htau}$ then $\PP{e} : \AA{\hG} \to \AA{\htau}$.
\refThm{static-correctness} immediately implies that the annotations
we obtain (statically) by provenance analysis conservatively
over-approximate the dependency-correct annotations we obtain
(dynamically) by provenance tracking provided the initial value $\hg$
matches $\AA{\hG}$.

We first establish that the static analysis is a conservative
extension of the ordinary type system:
\begin{lemma}\labelLem{conservativity-forward}
If $\wf{\Gamma}{e}{\tau}$ then for any $\hG \enriches \Gamma$ there exists a $\htau \enriches \tau$ such that $\wf{\hG}{e}{\htau}$.
\end{lemma}
\begin{proof}
  Structural induction on derivations; again the only interesting
  steps are those involving compatibility side-conditions;
  typically we only need to observe that if $\htau_1,\htau_2 \enriches
  \tau$ then $\htau_1 \cong \htau_2$, so $\htau_1 \sqcup \htau_2$
  exists and $\htau_1 \cong \htau_2 \cong \htau_1 \sqcup \htau_2$.
\end{proof}
\begin{lemma}\labelLem{conservativity-backward}
If $\wf{\hG}{e}{\htau}$ then $\wf{|\hG|}{e}{|\htau|}$.
\end{lemma}
\begin{proof}
  Straightforward induction on derivations; cases with compatibility side-conditions require observing that by definition $\htau_1 \cong \htau_2 \iff |\htau_1| = |\htau_2|$.
\end{proof}
\begin{lemma}\labelLem{enrichments-exist}
  Every context $\Gamma$ has at least one enrichment $\hG \enriches
  \Gamma$.
\end{lemma}
\begin{proof}
  Observe that any type can be lifted to an a-type by annotating each part of it with $\emptyset$.  An unannotated context $\Gamma$ can be lifted to a default annotated context $\hG$ by lifting each type.
\end{proof}
\begin{theorem}
  The judgment $\wf{\Gamma}{e}{\tau}$ is derivable if and only if for
  any $\hG$ enriching $\Gamma$, there exists a $\htau$ enriching
  $\tau$ such that $\wf{\hG}{e}{\htau}$ is derivable for some
  $\htau$ enriching $\tau$.
\end{theorem}
\begin{proof}
  For the forward direction, we use \refLem{conservativity-forward}.
  For the reverse direction, suppose the second part holds for a given
  $\Gamma,e,\tau$.  By \refLem{enrichments-exist}, we have
  $\wf{\hG}{e}{\htau}$ for some $\hG$ enriching $\Gamma$ and $\htau$
  enriching $\tau$.  Hence by \refLem{conservativity-backward}, we
  have $\wf{|\hG|}{e}{|\htau|}$, but clearly $|\hG| = \Gamma$ and
  $|\htau| = \tau$.
\end{proof}

We next establish useful properties of the a-value operations with
respect to the semantics of annotated types:

\begin{lemma}\label{lem:ops-well-typed}
For any $\Phi,\Psi,\Phi_1,\Phi_2,\htau,\htau_1,\htau_2$:
\begin{enumerate}
\item $\hop{+} : \AA{\kint^\Phi} \times \AA{\kint^\Psi} \to
  \AA{\kint^{\Phi \cup \Psi}}$.
\item $\hop{\sum} : \AA{\{\kint^\Phi\}^\Psi}\to \AA{\kint^{\Phi \cup
      \Psi}}$.
\item $\hat{\neg} : \AA{\kbool^\Phi} \to \AA{\kbool^\Phi}$.
\item $\hop{\andd} : \AA{\kbool^\Phi} \times \AA{\kbool^\Psi} \to
  \AA{\kbool^{\Phi \cup \Psi}}$.
\item $\hat{\pi_i} : \AA{(\htau_1 \times \htau_2)^\Phi} \to
  \AA{\htau_i^{+\Phi}}$ for any $i \in \{1,2\}$
\item $\hop{\eq} : \AA{\htau_1} \times \AA{\htau_2} \to
  \AA{\kbool^{\|\htau_1\| \cup\|\htau_2\|}}$.
\item If $\htau_1 \cong \htau_2$ then $\hat{\kcond} : \AA{\kbool^\Phi}
  \times \AA{\htau_1} \times \AA{\htau_2} \to \AA{(\htau_1 \sqcup
    \htau_2)^{+\Phi}}$
\item If $\htau_1 \cong \htau_2$ then $\hat{\cup} :
  \AA{\{\htau_1\}^{\Phi_1}} \times \AA{\{\htau_2\}^{\Phi_2}} \to
  \AA{\{\htau_1 \sqcup \htau_2\}^{\Phi_1\cup\Phi_2}}$.
\item If $\htau_1 \cong \htau_2$ then $\hat{-} :
  \AA{\{\htau_1\}^{\Phi_1}} \times \AA{\{\htau_2\}^{\Phi_2}} \to
  \AA{\{\htau_1\}^{\Phi_1 \cup\|\htau_1\| \cup\Phi_2 \cup\|\htau_2\|}}$.
\item $\hat{\bigcup} : \AA{\{\{\htau\}^\Psi\}^\Phi} \to
  \AA{\{\htau\}^{\Phi\cup\Psi}}$.
\end{enumerate}
\end{lemma}
\begin{proof}
  All of the properties are immediate from the definitions of the
  operations.
\end{proof}

\begin{theorem}\labelThm{static-correctness}
  If $\wf{\hG}{e}{\htau}$ then $\PP{e} : \AA{\hG} \to \AA{\htau}$.
\end{theorem}
\begin{proof}
  The proof is by induction on the structure of expressions (and the
  associated annotated derivations).  As before, many of the cases
  follow immediately by induction and appeals to
  Lemma~\ref{lem:ops-well-typed}.
  \begin{itemize}
  \item Case $e = x$: 
\[
\infer{\wf{\hG}{x}{\htau}}{x{:}\htau \in \hG} 
\]
Note that $\PP{x}\hg = \hg(x) \in \AA{\htau}$ since $\hg \in
\AA{\hG}$.
  \item Case $e = (\letin{x=e_1}{e_2})$: 
\[
\infer{\wf{\hG}{\letin{x=e_1}{e_2}}{\htau_2}}
{
\wf{\hG}{e_1}{\htau_1}
& 
\wf{\hG,x{:}\htau_1}{e_2}{\htau_2}
}
\]
By induction on the first subderivation, we have $\PP{e_1}\hg \in
\AA{\htau_1}$.  Hence $\hg[x:=\PP{e_1}\hg] \in \AA{\hG,x{:}\htau_1}$,
so by induction on the second subderivation, we have $\PP{e}\hg =
\PP{e_2}\hg([x:=\PP{e_1}\hg]) \in \AA{\htau_2}$
\item Case $e=(e_1,e_2)$:
\[
\infer{\wfeff{\hG}{(e_1,e_2)}{(\htau_1 \times \htau_2)}{\emptyset}}
{
\wf{\hG}{e_1}{\htau_1}
&
\wf{\hG}{e_2}{\htau_2}
}
\]
By induction, $\PP{e_i}\hg \in \AA{\htau_i}$.  Thus
$(\PP{e_1}\hg,\PP{e_2}\hg)^\emptyset \in \AA{(\htau_1 \times \htau_2)^\emptyset}$
\item Case $e =\{e'\}$: Similar to the case for pairing.
  \item Case $e =\{e_2 \mid x \in e_1\}$:
\[
\infer{\wfeff{\hG}{\{e_2 \mid x \in e_1\}}{\setof{\htau_2}}{\Phi_1}}
{
\wfeff{\hG}{e_1}{\setof{\htau_1} }{ \Phi_1} 
& 
\wf{\hG,x{:}\htau_1}{e_2}{\htau_2}
}
\]
Let $w^\Psi = \PP{e_1}\hg$; then by induction $w^\Psi \in
\AA{\{\htau_1\}^{\Phi_1}}$ and so $w \in \AA{\{\htau_1\}}$ and $\Psi
\subseteq \Phi_1$.  Hence for each $v \in w$, we have $v \in
\AA{\htau_1}$, so $\hg[x:=v] \in \AA{\hG,x{:}\htau_1}$.  Thus, for
each such $v$, by induction we have $\PP{e_2}(\hg[x:=v]) \in
\AA{\htau_2}$.  Moreover, $\PP{e}\hg = \{\PP{e_2}(\hg[x:=v]) \mid v
\hop{\in} w^\Psi\}= \{\PP{e_2}(\hg[x:=v]) \mid v \in w\}^{\Psi} \in
\AA{\{\htau_2\}^{\Phi_1}}$ since $\Psi \subseteq \Phi_1$.
  \end{itemize}
\end{proof}

%--------------------------------------------------------------------%
\section{Discussion}\labelSec{discussion}
%--------------------------------------------------------------------%

We chose to study provenance via the NRC because it is a clean and
system-independent core calculus similar to other functional
programming languages for which dependence analysis is
well-understood.  We believe our results can be specialized to common
database implementations and physical operators without much
difficulty.  We have not yet investigated scaling this approach to
large datasets or incorporating it into standard relational databases.

We have, however, implemented a prototype NRC interpreter that
performs ordinary typechecking and evaluation as well as provenance
tracking and analysis.  Our prototype currently displays the input and
output tables using HTML and uses embedded JavaScript code to
highlight backward slices, that is, the parts of the input on
which the selected part of the output may depend, according to the
analysis.  Similarly, the system displays the type information inferred
for the query and uses the results of static analysis to highlight
relevant parts of the input types for a selected part of the output
type.

% \todo[11]{Make online demo and discuss it further?  Say that we used
%   it to generate examples in this article.}

In the worst (albeit unusual) case, a part of the output could be
reported as depending on every part of the input, as a result of
spurious dependencies.  For example, this is the case for a query such
as $(R_1 - R_1) \cup \cdots \cup (R_n - R_n) \cup e$.  Of course, this
query is equivalent to $e$ and a good query optimizer will recognize
this.  However, for non-pathological queries encountered in practice,
our analysis appears to be reasonably accurate.  Even so, typically the
structure of the output depends on a large set of locations, such as
all of the fields in several columns in the input used in a selection
condition; individual fields in the output usually also depend on a smaller
number of places from which their values were computed or copied.
Thus, implementing provenance tracking in a large-scale
database may require developing more efficient representations
for large sets of annotations, especially the common case where a part
of the output depends on every value in a particular column.

The model we investigate in this article is similar to that of
\citeN{buneman08tods} in many respects.  There are two
salient differences.  The first difference is that
\citeN{buneman08tods} propagates annotations comprising single
(optional) input locations, whereas our approach propagates
annotations consisting of \emph{sets} of input locations.  The second
difference is that our approach provides a strong semantic guarantee
formulated in terms of dependence, whereas in contrast the semantics of
where-provenance in \citeN{buneman08tods} is an ad hoc syntactic
definition justified by a database-theoretic expressiveness result,
not a dependency property.

Of these differences, the second is more significant. Their results
characterize the possible where-provenance behavior of queries and
updates precisely, but tell us little about what might happen if the
input is changed.  Moreover, their expressiveness results have not yet
been extended to handle features such as primitive operations on data
values and aggregation.  To illustrate the distinction, observe that
in the example in \refFig{example}, the Name fields are copied from
the input to the output (thus, they have where-provenance in Buneman
et al.'s model) but the AvgMW fields are computed from several
sources, not copied (thus, they would have no where-provenance, even
though they depend on many parts of the input).  We believe the
approaches are complementary: each does something useful that the
other does not, and in general users may want both kinds of provenance
information to be available.

\citeN{buneman08tods} discuss implementing provenance tracking as a
source-to-source translation from NRC queries to NRC extended with a
new base type $\colorTy$.  The idea is to translate ordinary types to
types in which each subexpression is paired with an annotation of type
$\colorTy$, and translate provenance-tracking queries to ordinary NRC
queries over the annotated types that explicitly manage the
annotations.  This implementation approach has the potential advantage
that we can re-use existing query optimization techniques for NRC.  A
similar query-translation approach should be possible for dependency
provenance, by explicitly annotating each part of each value with a
set of annotations $\{\colorTy\}$ and using NRC set operations to
propagate colors.

% \todo[6]{Compare dependency-provenance with where-provenance.  Where is
%   always contained in dep, but we handle aggregation etc. better.}

Most database systems only implement SQL, which is less
expressive than NRC since it does not provide the ability to nest sets
as the field values of relations.  Nevertheless, it still may be
possible to support some annotation-propagation operations within
ordinary SQL databases.  Suppose we are interested in a particular
application in which annotations are
numerical timestamps or quality rankings that can be aggregated (e.g.
by taking the minimum or maximum).  In this case, we can propagate the
annotations from the source data to the results according to the
provenance semantics.  Simple SQL queries can easily be translated to
equivalent SQL queries that automatically aggregate annotations in
this way, using techniques similar to those used in the DBNotes
system~\cite{bhagwat05vldbj}.

For example, consider the query
\begin{verbatim}
SELECT A, SUM(B) FROM R GROUP BY A
\end{verbatim}
over relations $R: \{(A:\kint,B:\kint)\}$.  Suppose we have relations
$R: \{(A:\kint,A_q:\kint,B:\kint,B_q:\kint)\}$, in which each field
$A,B$ has an accompanying quality rating $A_q,B_q$.  Then we can
translate the above SQL query to
\begin{verbatim}
SELECT A, MIN(A_q), SUM(C), MIN(C_q) FROM R GROUP BY A
\end{verbatim}
to associate each value in the output with the minimum quality ranking
of the contributing fields---thus, data in the result with a high
quality ranking must depend only on high-quality data.  However,
performing this translation for general SQL queries appears
nontrivial.  It is well known that flat NRC queries whose input and
output types do not involve nested set types and that do not involve
grouping or aggregation and map sets of records to sets of records can
be translated back to SQL via a normalization
process~\cite{wong96jcss}, but it is apparently not well-understood
how to extract SQL queries from arbitrary NRC expressions involving
grouping and aggregation.

We can easily implement static provenance tracking for ordinary SQL
queries by translating them to NRC; this does not require changing the
database system in any way, since we do not need to execute the
queries.  Static provenance analysis is slightly more expensive
than ordinary typechecking, but since the overhead is proportional
only to the size of the schema and query, not the (usually much
larger) data, this overhead is minor.  Moreover, static analysis may
be useful in optimizing provenance tracking, for example by using the
results of static analysis to avoid tracking annotations that are
statically irrelevant to the output.

Consider for example the following scenario: After running a query,
the user identifies an error in the results, and requests a data slice
showing the input parts relevant to the error.  We can first provide
the results of static provenance analysis and show the user which
parts of the input database contain data that may have contributed to
the error.  In a typical relational database, this would narrow things
down to the level of database tables and columns, which may be enough
for the user to fix the problem.  In case this is not specific enough,
however, we can still employ the static analysis to speed
computing the dynamic provenance.  Using the static provenance
information, we know that only locations in the input data that
correspond to the static provenance of the output location of interest
can contribute to that output location.  Hence, if  we are only
interested in the dynamic provenance of a single output location,
we might avoid the overhead of dynamically tagging and tracking
provenance for parts of the input that we know cannot contribute to
the output part of interest.  We plan to investigate this potential
optimization technique in future work.

%--------------------------------------------------------------------%
\subsection{Comparison with slicing, information flow and other
  dependence analyses}
%--------------------------------------------------------------------%
The techniques in \refSec{provenance-tracking} and
\refSec{provenance-analysis} draw upon standard techniques in static
analysis~\cite{nielson05ppa,palsberg01paste}.  In particular, the idea
of instrumenting the semantics of programs with labels that capture
interesting dynamic properties is a well-known technique used in
control-flow analysis and information flow control.  Moreover, it
appears possible to cast our results in the \emph{abstract
  interpretation} framework~\cite{cousot77popl} that is widely used in
static analysis (see e.g.~\citeN[ch. 4]{nielson05ppa} for an
introduction).  Doing so would require adapting abstract
interpretation to handle collection types.  This does not appear
difficult but we preferred to keep the development in this article
elementary in order to remain accessible to nonspecialists.

Dependence tracking and analysis have been shown to be useful in many
contexts such as program slicing, information-flow security,
incremental update of computations, and
memoization and caching.  A great deal of work has been done on each
of these topics, which we cannot completely survey here.  We focus on
contrasting our work with the most closely related work in these areas.

%\todo[1]{More discussion of slicing and information flow}
%\todo[1,5]{How much is new here? Focus on collection types}
% \todo[8]{Mention program analysis inspiration and abstract interpretation
%   which we're not using}
 
In \emph{program slicing}~\cite{biswas97phd,field98ist,weiser81icse},
the goal is to identify a (small) set of program points whose
execution contributes to the value of an output variable (or other
observable behavior).  This is analogous to our approach to
provenance, except that provenance identifies relevant parts of the
\emph{input database}, not the \emph{program} (i.e.  query).
\citeN{cheney07debull} discusses the relationship between program
slicing and dependency analysis at a high level, complementing the
technical details presented in this article.

In computer security, it is often of interest to specify and enforce
\emph{information-flow policies}~\cite{sabelfeld03sac} that ensure
that information marked secret can only be read by privileged users,
and that privileged users cannot leak secret information by writing it
to public locations.  These properties are sometimes referred to as
\emph{secrecy} and \emph{integrity}, respectively.  Both can be
enforced using static (e.g.~\cite{myers99popl,volpano96jcs} or
dynamic (e.g. ~\cite{shroff08jcs,jia08icfp}) dependency tracking
techniques.  Our work is closely related to ideas in information flow
security, but our goal is not to prevent unauthorized disclosure but
instead to explicate the dependencies of the results of a query on its
inputs.  Nevertheless there are many interesting possible connections
that need to be explored, particularly in relating provenance to
dynamic information flow tracking~\cite{shroff08jcs} and
integrating provenance security policies with other access-control,
information-flow and audit policies~\cite{DBLP:conf/sp/SwamyCH08,jia08icfp}.

In contrast to most work on static analysis and information flow
security, we envision the instrumented semantics actually being used
to provide feedback to users, rather than only as the basis for
proving correctness of a static analysis or preventing security
vulnerabilities.  This makes our approach closest to (dynamic)
slicing.  The novelty of our approach with respect to slicing is that
we handle a purely functional, terminating database query language and
focus on calculating dependencies and slicing information having to do
with the (typically large) input data, not the (typically small)
query.  On the one hand the absence of side-effects, higher-order
functions, nontermination, or high-level programming constructs such
as objects and modules simplifies some technical matters considerably,
and enables more precise information to be tracked; on the other, the
presence of collection types and query language constructs leads to
new complications not handled in prior work on information flow,
slicing or static analysis.

In \emph{self-adjusting computation} \cite{acar08popl,acar09pepm},
support for efficient incremental recomputation is provided at a
language level.  Programs are executed using an instrumented semantics
that records their dynamic dependencies in a \emph{trace}.  Although
the first run of the program can be more expensive, subsequent changes
to the input can be propagated much more efficiently using the trace.
We are currently investigating further applications of ideas from
self-adjusting computation to provenance, particularly the use of
traces as explanations.

Dependency tracking is also important in \emph{memoization and
  caching} techniques~\cite{abadi96icfp,AcarBlHa03}.  For example,
\citeN{abadi96icfp} study an approach to caching the results of
function calls in a software configuration management system, based on
a label-propagating operational semantics.  \citeN{AcarBlHa03} develop
a language-based approach to memoizing and caching the results of
functional programs.  Our work differs from this work in that we
contemplate retaining dependency information as an aid to the end-user
of a (database) system, not just as an internal data structure used
for improving performance.

Our approach to provenance tracking based on dependency analysis has
been used in the Fable system~\cite{DBLP:conf/sp/SwamyCH08}.  In
this work provenance is one of a large class of security policies that
can be implemented using Fable, a dependently-typed language for
specifying security policies.
Subsequently,~\citeN{DBLP:conf/icfp/SwamyHB09} have explored a theory
of typed coercions that can be used to implement dependency
provenance.

\citeN{abadi99popl} argue that techniques such as slicing,
information-flow security, and other program analyses such as
binding-time analysis can be given a uniform treatment by translating
to a common \emph{Dependency Core Calculus}.  We believe provenance
may also fit into this picture, but in this article, we considered
both dynamic and static labeling, whereas the Dependency Core Calculus
only allows for static labels. Another difference is that the
Dependency Core Calculus is a higher-order, typed lambda-calculus
whereas here we have considered the first-order nested relational
calculus. It would be interesting to develop a common calculus that
can handle both static and dynamic dependence and both higher-order
functions and collections, particularly if dynamic information flow
and dynamic slicing could also be handled uniformly.

%\todo{Mention Hicks et al who have implemented variants of this}

%--------------------------------------------------------------------%
\section{Conclusions}\labelSec{concl}
%--------------------------------------------------------------------%

Provenance information that relates parts of the result of a database
query to relevant parts of the input is useful for many purposes,
including judging the reliability of information based on the relevant
sources and identifying parts of the database that may be responsible
for an error in the output of a query.  Although a number of
techniques based on this intuition have been proposed, some are ad hoc
while others have proven difficult to extend beyond simple conjunctive
queries to handle important features of real query languages such as
grouping, aggregation, negation and built-in operations.

We have argued that the notion of \emph{dependence}, familiar from
program slicing, information flow security, and other program
analyses, provides a solid semantic foundation for understanding
provenance for complex database queries.  In this article we
introduced a semantic characterization of \emph{dependency
  provenance}, showed that minimal dependency provenance is not
computable, and presented approximate tracking and analysis
techniques.  We have also discussed applications of dependency
provenance such as computing forward and backward data slices that
highlight dependencies between selected parts of the input or output.
We have implemented a small-scale prototype to gain a sense of the
usefulness and precision of the technique.

We believe there are many promising directions for future work,
including implementing efficient practical techniques for large-scale
database systems, identifying more sophisticated and useful dependency
properties, and studying dependency provenance in other settings such
as update languages and workflows.

%--------------------------------------------------------------------%
\subsection*{Acknowledgments} 
%--------------------------------------------------------------------%
We wish to thank Peter Buneman, Shirley Cohen and Stijn Vansummeren
for helpful discussions on this work.

%%% Local Variables: 
%%% mode: latex
%%% TeX-master: t
%%% End: 

% LocalWords:  Umut Acar Amal recomputation workflows timestamp metadata tuples
% LocalWords:  Abadi et al Ipso SQL monad  typechecked multisets subvalue iff
% LocalWords:  typechecking dom rclcrcl Buneman Vansummeren analyses pointwise
% LocalWords:  recolorings subterm sublanguages sublanguage coNP sem ann evy ok
% LocalWords:  Lampson Madnick polygen Cui Widom Khanna DBNotes hoc  SQL's ftdb
% LocalWords:  geospatial workflow TODO concl counterfactual enrichments tuple
% LocalWords:  noncomputable subtyping RDBMS DBMSs natively semirings semiring
% LocalWords:  memoizing annots renamings hrho memoization multiset NRC's versa
% LocalWords:  desugared datasets Biswas subterms lcl recordkeeping Stijn cy mw
% LocalWords:  thioredoxin flavodoxin ferredoxin ArgR CheW EnzReact PID thia IH
% LocalWords:  phos thi diphos acyl carboxylate ribose phosp ribulose gluc CoA
% LocalWords:  fruct panteth dephos AvgMW EnzymaticReaction ance proven EPSRC
% LocalWords:  abadi popl cccc ccc monads ringads rcl filenames monoid unlifted
% LocalWords:  equationally boolean conservativity cheney debull icfp ana icdt
% LocalWords:  Woodruff Stonebraker buneman subtrees nonemptiness subexpression
% LocalWords:  subexpressions de subderivation AcarBlHa woodruff subqueries dep
% LocalWords:  Javascript icde tods DAGs nielson ppa stateful nontermination

%%% Local Variables:
%%% TeX-master: "paper.tex"
%%% End:
% LocalWords:  JavaScript

\bibliographystyle{plainnat}
\bibliography{paper}

\end{document}